\documentclass{aa}
\usepackage{graphicx}
\usepackage{txfonts} 
\newcommand{\HI}{\ion{H}{i}}

\newcommand{\NI}{[\ion{N}{i}]$_{\lambda5199}$}
\newcommand{\NII}{[\ion{N}{ii}]$_{\lambda6583}$}

\newcommand{\OIII}{[\ion{O}{iii}]$_{\lambda5007}$}

\newcommand{\Msun}{$M_{\odot}$}
\newcommand{\agessp}{$t_{\rm SSP}$}
\newcommand{\metalssp}{[$Z$/H]$_{\rm SSP}$}
\newcommand{\abundssp}{[E/Fe]$_{\rm SSP}$}

\begin{document}
\title{Stellar populations, neutral hydrogen and ionised gas in field early-type galaxies}

%   \subtitle{put the subtitle here}

   \author{
           Paolo Serra \inst{1},
           Scott C. Trager \inst{1},
           Tom A. Oosterloo \inst{1,2},
           Raffaella Morganti \inst{1,2}
           }

%   \offprints{who to send offprints request to}

   \institute{
              (1) Kapteyn Astronomical Institute, University of Groningen, P.O. Box 800, NL-9700 AV Groningen, the Netherlands \\
%              \email{wuchterl@amok.ast.univie.ac.at}
%              \fnmsep\thanks{Just to show the usage of the elements in the author field}
%         \and
              (2) Netherlands Foundation for Research in Astronomy, Postbus 2, NL-7990 AA Dwingeloo, the Netherlands \\
%             \email{pserra@astro.rug.nl}
%             \thanks{The university of heaven temporarily does not accept e-mails}
             }

   \date{Received ...; accepted ...}

% \abstract{}{}{}{}{} 
% 5 {} token are mandatory
 
  \abstract
{}
{We present a study of the stellar populations of a sample of 39 local, field early-type galaxies whose \HI\ properties are known from interferometric data. Our aim is to understand whether stellar age and chemical composition depend on the \HI\ content of galaxies.  As a by-product of our analysis, we also study their ionised gas content and how it relates to the neutral hydrogen gas.}
{Stellar populations and ionised gas are studied from optical long-slit spectra. We determine stellar age, metallicity and alpha-to-iron ratio by analysing a set of Lick/IDS line-strength indices measured from the spectra after modelling and subtracting the ionised-gas emission.}
{We do not find any trend in the stellar populations parameters with $M$(\HI). However, we do find that, at stellar velocity dispersion $\sigma\leq 230$ km/s, 2/3 of the galaxies with less than $10^8$ \Msun\ of \HI\ are centrally rejuvenated, while none of the \HI-richer systems is. Furthermore, none of the more massive, $\sigma\geq 230$ km/s-objects is centrally rejuvenated independently on their \HI\ mass. Concerning the ionised gas, we detect emission in 60\% of the sample. This is generally extended and always carachterised by LINER-like emission-line ratios at any radius. We find that a large \HI\ mass is necessary (but not sufficient) for a galaxy to host bright ionised-gas emission.}
{A plausible interpretation of our results is that gas-rich mergers play a significant role in E/S0 formation, especially at lower $\sigma$. Within this picture, \HI-poor, centrally-rejuvenated objects could form in mergers where gas angular-momentum removal (and therefore inflow) is efficient; \HI-rich galaxies with no significant age gradients (but possibly uniformly young) could be formed in interactions characterised by high-angular momentum gas.}


   \authorrunning{Serra et al.}
   \titlerunning{Stellar populations, \HI\ and ionised gas in field E/S0's}
   \maketitle

\section{Introduction}

Most of the stellar mass of the local Universe is found in early-type (E and S0) galaxies (Baldry et al.\ 2004). Understanding the formation and evolution of these objects is therefore a key problem in modern astronomy. Historically, the two competing theories for E/S0 formation are the monolithic collapse from initial over-densities accompanied by early, efficient star formation (e.g., Eggen et al.\ 1962; Larson 1975), and the Toomre (1977) hypothesis that spheroidal galaxies form from mergers of disk systems. It is not yet clear which of them (or which of their combinations) matches better the observations. 

High-redshift surveys have made it possible to directly observe galaxy evolution over a large fraction of the cosmic history. An important result is that the stellar mass locked in the red-sequence (containing mostly spheroids, as opposed to the disk-dominated blue cloud; see Bell et al.\ 2004a) has roughly doubled since z$\sim$1 (e.g., Bell et al.\ 2004b; Faber et al.\ 2007). This result is incompatible with the monolithic picture, in which early-type galaxies are basically a static population.

Following Bell et al.\ (2004b) and Faber et al.\ (2007), the increase of the stellar mass in spheroids cannot be caused by star formation within the red sequence, as galaxies would become too blue. On the contrary, it must be caused by the migration of objects from the blue cloud to the red-sequence. Such colour evolution can occur by stellar ageing of blue galaxies provided that their star formation is quenched or reduced to very low levels (Harker et al.\ 2006). However, blue galaxies are mostly disks while red galaxies are mostly spheroids. Therefore, some morphological transformation must also occur.

Simulations show that disk-galaxy mergers can result in the formation of early-type systems (e.g., Toomre 1977; Mihos \& Hernquist 1994, 1996; Barnes \& Hernquist 1996). Thus, the above considerations make disk mergers a necessary ingredient for the formation of a large fraction of red-sequence objects; and the build-up of roughly half of the local E/S0's stellar mass since z$\sim$1 implies that many nearby early-type galaxies should show signatures of their recent merger origin. This is particularly true for galaxies fainter than $M_B=-21$, as Brown et al.\ (2007) have shown that most of the evolution takes place at or below the knee of the luminosity function.

Indeed, accurate inspection of E/S0's optical images often reveals low-surface-brightness fine structure (shells, ripples, plumes and others) in their stellar morphology, indicating recent dynamically-violent events (Malin \& Carter 1983; Schweizer \& Seitzer 1992). Furthermore, analysis of optical spectra and of UV photometry shows that a significant fraction of E/S0's hosted low-level star formation (a few percent in mass) within the past few Gyr (e.g., Trager et al.\ 2000; Yi et al.\ 2005). Such star formation could have been triggered by merging activity.

Within this picture, gas is likely to play a relevant role in shaping the local early-type galaxy population. For example, depending on the available gas mass and on the geometry of a merger, strong bursts of star formation can occur in the centre of the stellar remnant (e.g., Mihos \& Hernquist 1994, 1996; Kapferer et al.\ 2005; Cox et al.\ 2006; Di Matteo et al.\ 2007), affecting its average age and chemical composition. Furthermore, high angular-momentum tidal-tail gas could survive the merging process and settle on stable orbits around the newly-formed galaxy (Barnes 2002). It is possible that in some cases this leads to the formation of a new stellar disk, while in general the result will be an \HI-rich spheroid. Finally, the stellar orbital structure of the remnant would be affected by the formation of axisymmetric gaseous systems within the optical body (Naab et al.\ 2006). The last point might be important for the emergence of disky, fast-rotating galaxies as opposed to boxy slow rotators (as proposed by Bender et al.\ 1992; see also Emsellem et al.\ 2007). Significant rotational support is indeed more common at lower stellar masses, which is consistent with Brown et al.\ (2007) claim that the evolution of E/S0's since z$\sim$1 concerns mostly smaller galaxies. All this implies that, within the merger picture, gas is intimately related to the formation and evolution of E/S0's stellar body. It is therefore natural to think that galaxy gas properties might be related to optical ones. The present paper is mostly concerned with the exploration of such connection.

\vspace{0.5cm}

Observationally, early-type galaxies have long been considered as very gas-poor objects. In fact, the detection of gas in E/S0's has mostly been hampered by its very low column density (hence surface brightness). We now know that E/S0 galaxies host a multi-phase gas component going from hot, X-ray-emitting gaseous halos (Forman et al.\ 1985; Fabbiano et al.\ 1992; O'Sullivan et al.\ 2001) to central warm-gas distributions (see Goudfrooij 1999 review) and  cold, neutral-hydrogen structures (see the review of Sadler et al.\ 2002; and more recent work by Morganti et al.\ 2006; Helmboldt 2007; Oosterloo et al.\ 2007).

In detail, $\sim$50-75\% of E/S0's host ionised gas commonly detected via its optical emission lines (e.g., Caldwell 1984; Phillips et al.\ 1986; Sarzi et al.\ 2006; Yan et al.\ 2006). This emission can seldom be explained in terms of on-going star formation, most of galaxies showing instead line-ratios typical of LINERs (Ho et al.\ 1997; Filippenko 2003; Yan et al.\ 2006). Furthermore, the ionised gas appears to be mostly diffuse. For example, studying the SAURON sample of E/S0's Sarzi et al.\ (2006) find warm gas in a variety of extended configurations generally kinematically decoupled from the stellar component (but less so in flat, fast-rotating galaxies).

Many studies of the colder neutral gas component of individual systems and samples of galaxies show that early-type galaxies host up to a few times 10$^{10}$ \Msun\ of \HI\ distributed out to many tens of kpc from the stellar body (Schiminovich et al.\ 1994, 1995, 1997; V\'{e}ron-Cetty et al.\ 1995; Morganti et al.\ 1997, 2006; Balcells et al.\ 2001; Oosterloo et al.\ 2002, 2007; Helmboldt 2007). Typical column densities are below 10$^{20}$ cm$^{-2}$ which makes it unlikely for the gas to host large-scale star formation. Importantly, unlike what is seen in spiral galaxies, the \HI\ mass does not correlate with the optical luminosity. This has been interpreted as a sign of external origin of the gas. Finally, we find a wide variety of morphologies and kinematics, ranging from long-lived, extended rotating disks or rings to unsettled clumps or tails of \HI.

\begin{table*}
\begin{center}
\caption{Sample galaxy properties}
\label{sampleprop}
\begin{tabular}{lcccrrrrrc}
\hline
\hline
\noalign{\smallskip}
\multicolumn{1}{c}{Galaxy}      & Type   & $d$  & $M_B$ & \multicolumn{1}{c}{$R_e$} & \multicolumn{1}{c}{$\sigma_{R_e/16}$} & \multicolumn{1}{c}{$S$(\HI)}         & \multicolumn{1}{c}{$M$(\HI)}     & \multicolumn{1}{c}{$\epsilon$} & References \\
            &        & (Mpc)&            & \multicolumn{1}{c}{(arcsec)}          & \multicolumn{1}{c}{(km/s)}          & \multicolumn{1}{c}{(Jy km s$^{-1}$)} &\multicolumn{1}{c}{(10$^9$\Msun)} & \multicolumn{1}{c}{(Mpc)}      &            \\
\multicolumn{1}{c}{(1)}         & (2)    & (3)  & (4) & \multicolumn{1}{c}{(5)}             & \multicolumn{1}{c}{(6)}              & \multicolumn{1}{c}{(7)}          & \multicolumn{1}{c}{(8)}        & \multicolumn{1}{c}{(9)}        &            \\
\noalign{\smallskip} \hline \noalign{\smallskip}
\noalign{\smallskip} \noalign{\smallskip}
\multicolumn{10}{c}{\emph{NTT/EMMI spectroscopy}}\\
ESO 092-21	& E-S0	 & 30.6 &   -19.39	 & 13 & 86			   & 19.2             & 4.2	         & 1.05    & b,d,e\\
ESO 140-31	& E		 & 47.2 &   -20.05	 &  7 & 163  		   & 7.7              & 4.0		     & 0.77    & b,d,e\\
ESO 381-47	& S0	 & 63.7 &   -19.76	 & 22 & 188  		   & 7.1              & 6.7		     & 1.10    & b,d,e\\
 IC 4200	& S0	 & 63.7 &   -21.48	 & 12 & 268  		   & 8.7              & 8.2		     & 1.49    & b,d,e\\
 IC 4889	& S0	 & 29.2 &   -20.53	 & 12 & 176  		   & 7.1              & 1.4		     & 2.21    & a,c,e\\
NGC 1490	& E		 & 74.8 &   -21.13	 & 18 & 311  		   & 5.7              & 7.4		     & 3.32    & b,d,e\\
NGC 1947	& S0	 & 14.3 &   -19.39	 & 22 & 150  		   & 3.0              & 0.14		 & 1.16    & a,d,f\\
NGC 2434	& E		 & 21.6 &   -20.41	 & 15 & 235  		   & -                & $<$0.1	     & $>$1.89 & a,c,e\\
NGC 2904	& E-S0	 & 23.6 &   -18.93	 & 12 & 238  		   & -                & $<$0.2		 & 1.03    & b,c,e\\
NGC 3108	& S0-a	 & 40.6 &   -20.65	 & 19 & 233  		   & 6.9              & 2.7		     & 0.37    & b,d,f\\
\noalign{\smallskip} \noalign{\smallskip}
\multicolumn{10}{c}{\emph{WHT/ISIS spectroscopy}}\\
NGC 0596	& E		 & 21.8 &   -20.03	 & 26 & 170   		   & -                & $<$0.02 	 & 0.24    & b,c,g\\
NGC 0636	& E		 & 29.8 &   -20.16	 & 22 & 184   		   & -                & $<$0.04	     & 1.02    & a,c,g\\
NGC 1426	& E		 & 24.1 &   -19.71	 & 21 & 157  		   & -                & $<$0.04 	 & 0.25    & b,c,h\\
NGC 1439	& E		 & 26.7 &   -19.99	 & 22 & 156  		   & -                & $<$0.04      & 0.43    & b,c,h\\
NGC 2300	& E 	 & 26.4 &   -20.38	 & 24 & 291  		   & -                & $<$0.03 	 & 0.49    & a,d,g\\
NGC 2534	& E		 & 48.3 &   -19.99	 & 18 & 165  		   & 1.4              & 0.76		 & 2.10    & b,d,h\\
NGC 2549	& S0	 & 12.6 &   -18.70	 & 25 & 150   		   & -                & $<$0.002	 & 1.01    & a,c,i\\
NGC 2768	& S0	 & 22.4 &   -21.15	 & 27 & 208  		   & 1.5              & 0.18		 & 1.28    & a,c,i\\
NGC 2810	& E		 & 50.8 &   -20.57	 & 20 & 248  		   & 1.2              & 0.72		 & 0.27    & b,d,h\\
NGC 3193	& E		 & 34.0 &   -20.80	 & 27 & 213   		   & -                & $<$0.03 	 & 0.53    & a,c,j\\
NGC 3610	& E-S0	 & 21.4 &   -20.10	 & 19 & 174   		   & -                & $<$0.009 	 & 0.39    & a,c,k\\
NGC 3640	& E		 & 27.0 &   -21.04	 & 22 & 203   		   & -                & $<$0.02 	 & 0.39    & b,c,k\\
NGC 3998    & S0	 & 14.1 &   -19.43	 & 23 & 300  		   & 6.4              & 0.30		 & 0.49    & a,c,m\\
NGC 4026	& S0	 & 13.6 &   -19.08	 & 40 & 194  		   & 51.3             & 2.2		     & 0.38    & a,c,n\\
NGC 4125	& E-S0	 & 23.9 &   -21.36	 & 26 & 248  		   & 0.2              & 0.027		 & 0.76    & a,c,o\\
NGC 4278	& E		 & 16.1 &   -20.13	 & 26 & 283  		   & 11.4             & 0.69		 & 0.23    & a,c,i\\
NGC 4406	& E		 & 17.1 &   -21.46	 & 27 & 262  		   & 1.22             & 0.083		 & 0.59    & b,c,p\\
NGC 4472	& E		 & 16.3 &   -21.91	 & 25 & 306  		   & 0.73             & 0.045		 & 0.21    & b,c,q\\
NGC 5018	& S0	 & 46.1 &   -22.09	 & 27 & 221  		   & 1.9              & 0.94	     & 0.37    & a,d,r\\
NGC 5173	& E		 & 38.4 &   -19.69	 & 20 & 118   		   & 4.7              & 1.6		     & 0.23    & b,d,s\\
NGC 5322	& E-S0	 & 31.2 &   -21.51	 & 21 & 251  		   & -                & $<$0.02 	 & 0.45    & a,c,k\\
NGC 5903	& E-S0	 & 33.9 &   -21.13	 & 27 & 225  		   & 3.4              & 0.91		 & 0.21    & a,c,t\\
NGC 7052    & E		 & 65.5 &   -20.97	 & 46 & 327  		   & -                & $<$0.065     & 3.52    & b,d,u\\
NGC 7332	& S0	 & 23.0 &   -20.01	 & 26 & 147  		   & 0.05             & 0.006		 & $>$2.01 & a,c,i\\
NGC 7457	& S0	 & 13.2 &   -18.99	 & 30 & 76  		   & -                & $<$0.002	 & $>$1.16 & a,c,i\\
NGC 7585	& S0-a	 & 46.3 &   -21.17	 & 25 & 217  		   & -                & $<$0.1  	 & 1.70    & a,d,g\\
NGC 7600	& E-S0	 & 45.1 &   -20.55	 & 28 & 193  		   & -                & $<$0.09 	 & 3.22    & a,d,g\\
NGC 7619	& E		 & 53.0 &   -21.94	 & 28 & 334  		   & 0.04             & 0.026		 & 0.15    & a,c,h\\
NGC 7626	& E		 & 44.3 &   -21.44	 & 30 & 288  		   & -                & $<$0.06 	 & 0.14    & a,d,h\\
\noalign{\smallskip}
\hline
\end{tabular}
\end{center}

(1) Galaxy identifier. 
(2) Morphological classification from: ($a$) Sandage \& Bedke (1994); ($b$) de Vaucouleurs et al.\ (1991). 
(3) Distance from: ($c$) Tonry et al.\ (2001) (surface-brightness fluctuation); ($d$) NED Hubble-flow velocity corrected for Virgo, Great Attractor and Shapley Supercluster infall ($h=0.7$). 
(4) Blue absolute magnitude from HyperLeda (Paturel et al. 2003) corrected apparent magnitude adopting distance $d$. 
(5) Effective radius $R_e$ estimated from the long-slit spectra presented in this paper (see Sec.\ref{obs}). 
(6) Stellar velocity dispersion within the inner $R_e$/16. $\sigma_{R_e/16}$ is the average of the values obtained along two perpendicular slit positions. 
(7) Integrated \HI\ flux from: ($e$) Oosterloo et al.\ (2007); ($f$) Oosterloo et al.\ (2002); ($g$) Sansom et al.\ (2000); ($h$) Dijkstra (1999); ($i$) Morganti et al.\ (2006); ($j$) Williams et al.\ (1991); ($k$) Hibbard \& Sansom (2003); ($l$) Balcells et al.\ (2001); ($m$) Knapp et al.\ (1985); ($n$) Verheijen (priv.com.); ($o$) Rupen et al.\ (2001); ($p$) Arecibo ALFALFA archive, Giovannelli et al.\ (2007); ($q$) McNamara et al.\ (1994); ($r$) Kim et al.\ (1988); ($s$) Knapp \& Raimond (1984); ($t$) Appleton et al.\ (1990); ($u$) Emonts (2006). 
(8) $M$(\HI)=2.36 10$^{5}$ ($d$/Mpc)$^2$ $S$(\HI)/(Jy km s$^{-1}$) \Msun. Upper limits are obtained by scaling the ones given in the references above to the distances adopted here and to the criterion of having a 3$\sigma$ signal over 200 km/s; or by applying this same criterion to the noise reported in the reference. 
(9) Projected distance from NED 4th closest neighbour brighter than $M_B$=--18 (HyperLeda corrected absolute magnitude) and within $\pm$500 km/s (using NED Hubble-flow fully-corrected velocities). $\epsilon$ is transformed from arcmin to Mpc adopting the scale at distance $d$.
\end{table*}

Some attempt to explore the connection between galaxy optical and gas properties has already been made. For example, van Gorkom \& Schiminovich (1997) showed that early-type galaxies with morphological fine structure are more likely to contain \HI\ gas. Along with this, Emsellem et al.\ (2007) found that systems with bright ionised-gas emission lines are fast rotators (i.e., the specific stellar angular-momentum is large). On the other hand, Morganti et al.\ (2006) found no relation between the stellar age and the \HI\ mass of a small sub-sample of the SAURON sample of E/S0's, while Graves et al.\ (2007) reported that LINER-like red galaxies within SDSS are statistically younger than equal-mass quiescent systems.

In the present work we investigate the relation between \HI\ and stellar content of E/S0's by means of radio and optical observations of a sample of $\sim$40 nearby galaxies. In particular, we want to answer one question: do the stellar age and chemical composition of early-type galaxies depend on their neutral-gas content? We find that, globally, they do not. However, our analysis suggests that, at velocity dispersion below 230 km/s, \HI-poor galaxies can host a central stellar population $younger$ than the underlying one; on the contrary, \HI-rich systems never do. Furthermore, \HI-poor galaxies with a central drop in age are characterised by a stronger central increase in metallicity than other objects.

As a by-product of our analysis we also study the ionised-gas content of the galaxies in the sample. We find that 60\% of them host (mostly diffuse) ionised gas with line-ratios typical of LINERs at any radius. Furthermore, a large $M$(\HI) appears $necessary$ to have strong ionised-gas emission.

In what follows we present the observed sample (Sec.2), optical observations, data reduction and analysis (Sec.3) and main results (Sec.4). We then discuss these results (Sec.5) and draw some conclusions.

\begin{figure}
\includegraphics[angle=270,width=8cm]{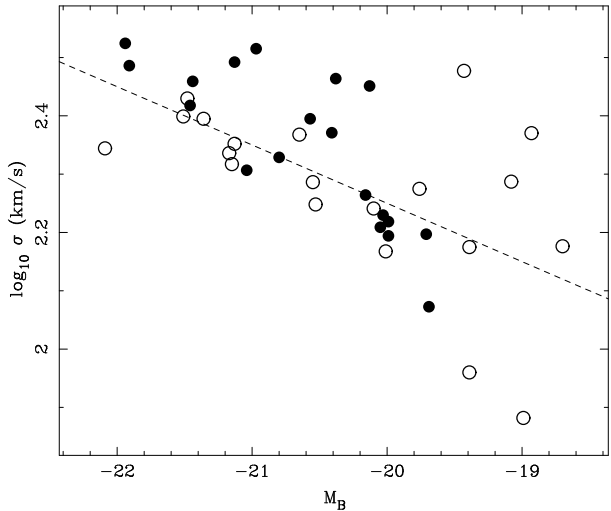}
\caption{Stellar velocity dispersion vs. absolute blue magnitude for ellipticals (filled circles) and lenticulars (open circles) in the sample. The dashed line represents the Faber-Jackson relation.}
\label{FJ}
\end{figure}

\section{The sample}
\label{sample}

In order to test whether there is a relation between stellar populations and \HI\ in E/S0's we select a sample of 39 field galaxies classified as early-type and covering a range in $M$(\HI). The sample contains \HI-rich and \HI-poor galaxies, i.e., systems that previous \emph{interferometric} 21-cm observations showed respectively to host and not to host \HI. Table \ref{sampleprop} summarises the main galaxy properties and how they are derived.

Fig.\ref{FJ} shows the galaxy distribution on the $M_B$-$log_{10}\sigma$ plane. Ellipticals and lenticulars are plotted as filled and open circles respectively. The dashed line corresponds to the Faber-Jackson relation $L_B\propto\sigma^4$ (Faber \& Jackson 1976), which galaxies in the sample follow reasonably well.

\begin{figure}
\includegraphics[width=8cm]{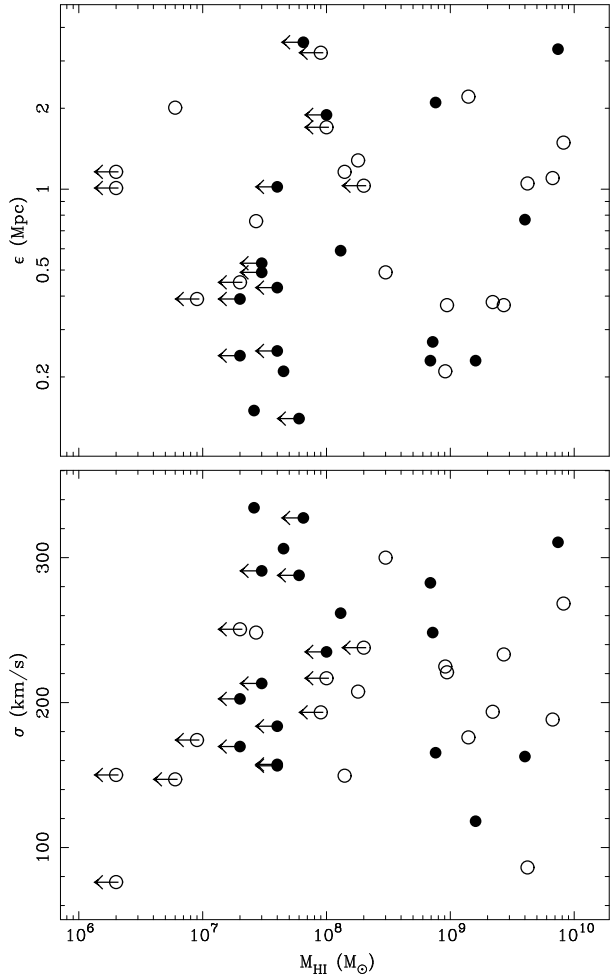}
\caption{Stellar velocity dispersion and $\epsilon$ (environment indicator) plotted against the \HI\ mass for ellipticals (filled circles) and lenticulars (open circles) in the sample. See Table \ref{sampleprop} for the definition of $\epsilon$. Arrows indicate \HI\ non-detections, for which the upper limit on $M$(\HI) is plotted.}
\label{sistematics}
\end{figure}

We note that galaxies are chosen among the relatively few objects with available 21-cm interferometric data before 2004. Such radio observations were carried out by different authors with different telescopes and on targets selected according to different criteria. Therefore, our sample is not representative of the local population of field early-type galaxies. On the other hand, it is suitable to test whether other E/S0 properties (e.g., stellar populations) depend on the \HI-gas content.

As we are looking for trends with $M$(\HI), it is important that there are no systematics in the sample. Fig.\ref{sistematics} shows the distribution of \HI\ masses with respect to $\sigma$ and $\epsilon$. The latter characterises the environment with the distance from the fourth closest neighbour (see Table \ref{sampleprop} for details). \HI-poor systems are highlighted with an arrow indicating that their $M$(\HI) value is an upper limit. Fig.\ref{sistematics} shows that there is no strong correlation between either \HI\ mass and stellar velocity dispersion or \HI\ mass and environment. Furthermore, the distributions of ellipticals and lenticulars are comparable. The only possible exception are the 4 \HI-poorest galaxies, which are all low-$\sigma$ S0's.

\begin{table}
\begin{center}
\caption{Long-slit spectroscopic observations}
\label{longslit}
\begin{tabular}{l c c}
\hline
\hline
\noalign{\smallskip}
                       & NTT/EMMI    & WHT/ISIS    \\
\noalign{\smallskip} \hline \noalign{\smallskip}
Observing runs         & 12/07/04    & 05-06/04/06 \\
                       & 04-07/02/05 & 13-14/10/06 \\
Arm                    & Red         & Blue        \\
Spectral range (\AA)   & 4000-7000   & 3700-7100   \\
Dispersion (\AA/pixel) & 1.66        & 1.7         \\
Resolution (km/s)      & 150         & 130         \\
Scale (arcsec/pixel)   & 0.33        & 0.4         \\
Slit length (arcmin)   & 8           & 4           \\
Slit width (arcsec)    & 1           & 1           \\
\noalign{\smallskip}
\hline
\end{tabular}
\end{center}
\end{table}

\section{Observations, data reduction and analysis}
\label{obs}

We observe all galaxies with long-slit optical spectroscopy. The sample contains objects in both hemispheres. Northern and southern galaxies are observed respectively with WHT/ISIS in La Palma, and NTT/EMMI at La Silla. Table \ref{longslit} summarises the instrumental set-up for both sets of observations.

We observe each galaxy along two perpendicular slit directions aligned to the optical major and minor axis. We choose exposure times that give $S/N$$\sim$100/\AA\ at 5000 \AA\ within the inner 1/8 of the effective radius in order to accurately determine stellar populations of galaxies. The seeing during the runs was 0.9 arcsec in the worst case. We reduce the data in the standard way with a variety of suitable programs written by Daniel Kelson, SCT and PS in the Python programming language (see Serra et al.\ 2006 for more details). The wavelength calibration rms never exceeds 0.1 \AA.

We re-bin the 2D spectra to a 1-arcsec spatial scale (equal to the slit width). To extract aperture-equivalent 1D spectra, we scale up the counts of all rows (but the central one) by their distance from the galaxy centre (Gonz\'{a}lez 1993). This allows us to calculate for each galaxy and along each slit position-angle the effective radius $R_e$ as the radius containing half of the total light. We note that given the depth of our long-slit observations this estimate is likely to be smaller than the true effective radius which would be measured from optical imaging. Instead, $R_e$ should be regarded as the scale of the central light distribution and is used in order to analyse the spatial variation of galaxy spectral properties in a consistent way across the sample. Values of $R_e$ are listed in Table \ref{sampleprop}.

We extract for each galaxy along each slit position angle aperture-equivalent 1D spectra using aperture radii of $R_e$/16 and $R_e$/2. Note that in doing so we adopt a minimum aperture radius of 0.75 arcsec to avoid seeing problems. In one case (ESO 140-31) this causes the smallest aperture to have radius $R_e$/9 rather than $R_e$/16. The spectra are then flux calibrated and smoothed to a velocity dispersion independent on wavelength and equal to 150 and 130 km/s for EMMI and ISIS spectra respectively. The resulting spectra (two per galaxy per aperture along the two perpendicular slit alignments) are then analysed as described below.

\subsection{Separating stellar spectrum from ionised-gas emission lines}
\label{gandalf}

As discussed above, early-type galaxies often show ionised-gas emission lines superimposed on the stellar spectrum. In order to properly study stellar populations it is necessary to first remove such contamination, which can then be analysed separately. We do this by means of GANDALF, a program which fits the best combination of stellar spectral templates and Gaussian emission lines to a given spectrum (for details, see Sarzi et al.\ 2006). We use GANDALF to derive (for each galaxy, slit position-angle and aperture) the purely-stellar spectrum by subtracting the best-fitting emission-line spectrum from the observed one.

In this work we fit MILES single-burst stellar population (SSP) models (Vazdekis et al. 2008, in prep.) and adopt a 11$^{\rm th}$-order multiplicative polynomial to match the stellar continuum over the entire observed wavelength range. GANDALF allows the user to fit any set of emission lines with various constraint on their relative kinematics and flux. As a general scheme, we force all forbidden lines to have the kinematics of \NII\ and all recombination lines to have the kinematics of H$\alpha$. However, in case of very weak emission, we fit first \OIII\ (easier to detect) and then tie the kinematics of all other lines to it.

We subtract the emission model-spectrum from the observed one (i.e., stellar$+$gas) after applying a line-detection criterion. Namely, we require $A/N\geq3$ for a line to be detected, where $A$ is the amplitude of the Gaussian fit to the line and $N$ is the median spectrum noise within a narrow wavelength region centred on the line. In fact, $N$ is scaled up by the $\chi^2$ of the fit to avoid large residuals (with respect to the spectrum noise) being detected as gas emission in cases of poor fitting. For each Balmer emission-line we relax the detection criterion to $A/N\geq2$ if the lower-order line is detected.

We also allow any line with $A/N\geq5$ to be kinematically free; i.e., we fit its kinematics independently to reduce line-subtraction residuals. This is important when the emission fills in a significant fraction of (or an entire) stellar absorption line (e.g., H$\beta$). We have tested that in these cases the gas kinematics derived from different lines are consistent. We have also tested that the flux ratios of H$\beta$ to H$\alpha$, H$\gamma$ and H$\delta$ are compatible with Case B recombination (Osterbrock \& Ferland 2005). Typical $\chi^2$ is 1 and $\chi^2\sim2$ is the worst case.

In two cases, NGC 3998 and NGC 7052, the ionised gas shows double Gaussian profiles, causing the ionised-gas removal to be of poor quality. While the estimate on the emission line flux of these two galaxies is still reasonably accurate, we do not analyse their stellar populations as indices affected by gas emission (see below) are likely to contain large errors.

Summarising, GANDALF allows us to obtain for each galaxy, aperture and slit position-angle: the purely-stellar spectrum as the observed spectrum minus the best-fitting emission spectrum; a set of detected emission lines and their flux; and the line-of-sight velocity distribution of stellar emission and emission lines. For each galaxy and aperture, the flux of a given line is taken as the average of the fluxes measured along the two perpendicular slit positions. Table \ref{t3} gives the flux of H$\beta$, \OIII, H$\alpha$ and \NII\ emission for all galaxies and apertures.

\subsection{Lick/IDS line-strength indices and derivation of \agessp, \metalssp\ and \abundssp}
\label{lick2ssp}

\begin{figure}
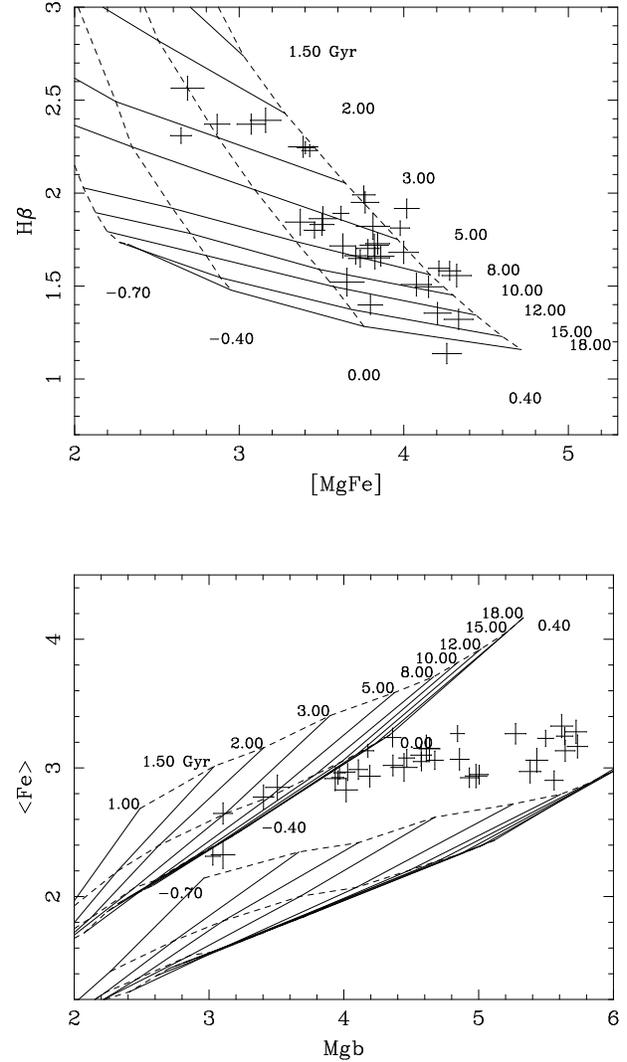

\includegraphics[angle=270,width=8cm]{MgFeHb.eps}

\vspace{1cm}
\includegraphics[angle=270,width=8cm]{MgFe.eps}
\caption{Line-strength indices measured within the $R_e$/16 aperture and plotted over BC03 model grids. Note that $<$Fe$>$=(Fe5270+Fe5335)/2 and [MgFe]=(Mg$b<$Fe$>$)$^{1/2}$. Error bars are 1 standard deviation estimated from the spectrum noise only. \emph{Top panel}: Only the [E/Fe]=0 grid is plotted. Solid lines have constant age while dashed lines have constant metallicity. Age and metallicity values label the corresponding grid lines. \emph{Bottom panel}: top and bottom grid have [E/Fe]=0 and +0.3 respectively. Grid lines are labelled with the corresponding age and metallicity only for the [E/Fe]=0 grid.}
\label{grid}
\end{figure}

We analyse the stellar populations of our galaxies by measuring Lick/IDS line-strength indices (Worthey et al.\ 1994; Worthey \& Ottaviani 1997) from the gas-free spectra, and comparing the result to model predictions as a function of stellar age and chemical composition. In order to correctly perform the comparison, we measure the indices after having broadened the observed spectra to match the Lick/IDS resolution. We then correct the measurements for the effect of velocity dispersion. For this purpose, we calculate for each spectrum and index a multiplicative correction as the factor that transforms the index measured from the best-fitting stellar spectrum to the one measured from its unbroadened ($\sigma_{stars}=0$ km/s) counterpart. Finally, we bring the line-strength indices onto the Lick/IDS system. We do this by means of MILES-library calibration stars (S\'{a}nchez-Bl\'{a}zquez et al.\ 2006) that we observed during each observing run. Details on this calibration can be found in Appendix A. Here we would like to stress that the entire sample is calibrated by means of MILES stars, so that all galaxies are on the same index system.

\begin{figure}
\includegraphics[angle=270,width=8cm]{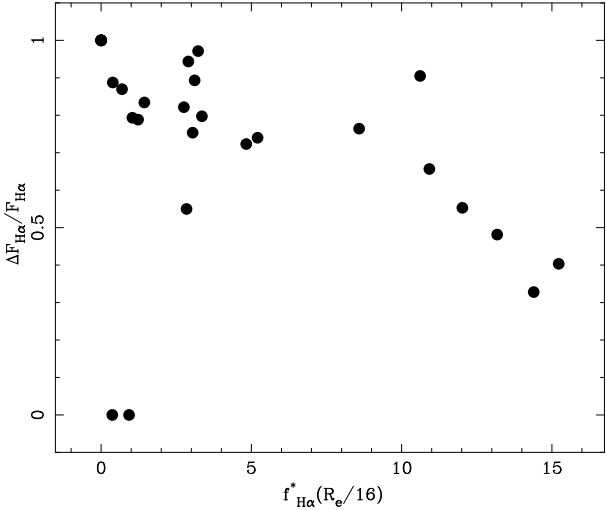}
\caption{Ionised gas detected outside $R_e$/16 relative to the total plotted against the specific gas content within $R_e$/16 (see text). $\Delta F_{H\alpha}/F_{H\alpha}=0$ or 1 means that all the gas is detected inside or outside $R_e$/16 respectively.}
\label{gasext}
\end{figure}

\begin{figure}
\includegraphics[width=8cm]{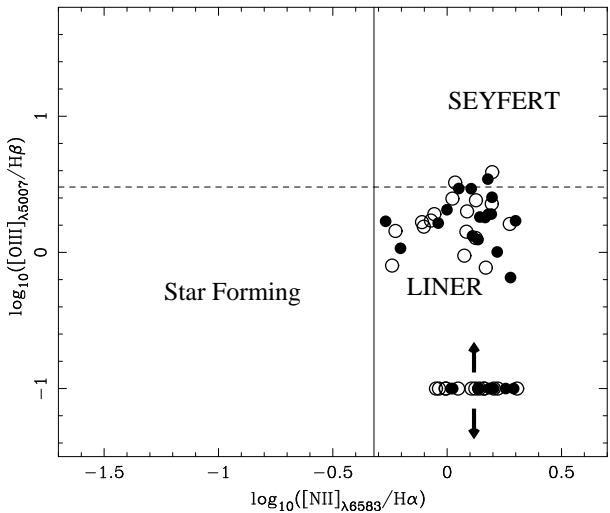}
\caption{Emission-line flux ratio diagram. Labels and dashed lines indicate schematically which region of the diagram corresponds to ionisation typical of star-forming regions, LINER and Seyfert galaxies, following Kauffmann et al.\ (2003). Filled and open circles correspond to $R_e$/16 and $R_e$/2 apertures respectively. All measured points lie within the LINER region. Points at \OIII/H$\beta$=0.1 are not detected in the lines \OIII and/or H$\beta$, while they are in \NII and H$\alpha$. These points could lie anywhere along the vertical axis.}
\label{bpt}
\end{figure}

We combine the line-strength indices along the two perpendicular slit positions for each galaxy and aperture. We then compare the measured value of H$\beta$, Mg$b$, Fe5270 and Fe5335 to the values predicted by Bruzual \& Charlot (2003, BC03) SSP models as a function of age $t$ and metallicity [Z/H]. In doing the comparison we take account of non-solar alpha-to-iron abundance ratios by parameterising them with the quantity [E/Fe] introduced by Trager et al.\ (2000a). For this purpose, we use Lee et al.\ (in prep.) index responses to [E/Fe] variations to modify BC03 index values (see Trager, Faber \& Dressler 2008, submitted, for more details).

Fig.\ref{grid} shows the data-to-model comparison by plotting the measured galaxy indices (for the $R_e$/16 aperture only) and  the models on two projections of the hyperspace defined by H$\beta$, Mg$b$, Fe5270 and Fe5335.  Comparison to SSP models allows us to derive the best-matching SSP-equivalent age \agessp, metallicity \metalssp\ and abundance ratio \abundssp. For each spectrum, these are calculated with the aid of 10000 Monte Carlo realisations of the four measured indices whose standard deviation is estimated from the spectrum noise. Each of \agessp, \metalssp\ and \abundssp\ is taken as the peak value of its probability distribution marginalised with respect to the other two parameters. The errors are defined to contain 68\% of the probability integral (see Trager, Faber \& Dressler 2007 for more details). Off-grid points are fitted by linearly extrapolating the models. Given the regular behaviour of the model grids in Fig.\ref{grid}, this treatment is reasonable in the proximity of the grid boundaries . Tables \ref{t4} and \ref{t5} list the index values and derived stellar parameters for all galaxies and apertures.

\section{Results}
\label{result}

The data analysis described above gives us for each galaxy and over each of the two apertures of radius $R_e/16$ and $R_e/2$: the stellar population parameters \agessp, \metalssp, \abundssp\ and  a set of emission-line fluxes relative to all detected lines. Furthermore,  we know for all galaxies $M$(\HI) or its upper limit (see Table \ref{sampleprop}). This section presents the results derived with respect to these quantities.

\subsection{Ionised gas: properties and connection to the \HI\ content}
\label{fstardefinition}

We detect ionised gas in 60\% of the sample. This is consistent with previous estimates of the fraction of E/S0's containg warm gas (e.g., Sarzi et al.\ 2006, Yan et al.\ 2006) even though our sample is not meant to be representative of the local early-type population.

\begin{figure}
\includegraphics[angle=270,width=8cm]{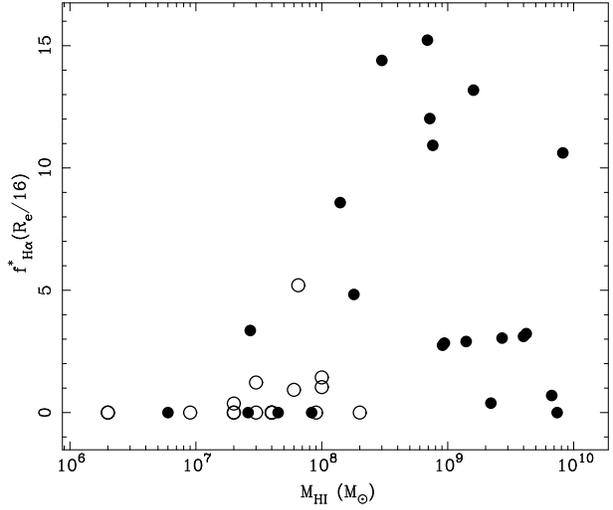}
\caption{Specific H$\alpha$ emission (defined in the text) plotted against the \HI\ mass. Filled and open circles represent \HI\ detections and upper limits respectively.}
\label{hahi}
\end{figure}

\begin{figure}
\includegraphics[angle=0,width=8cm]{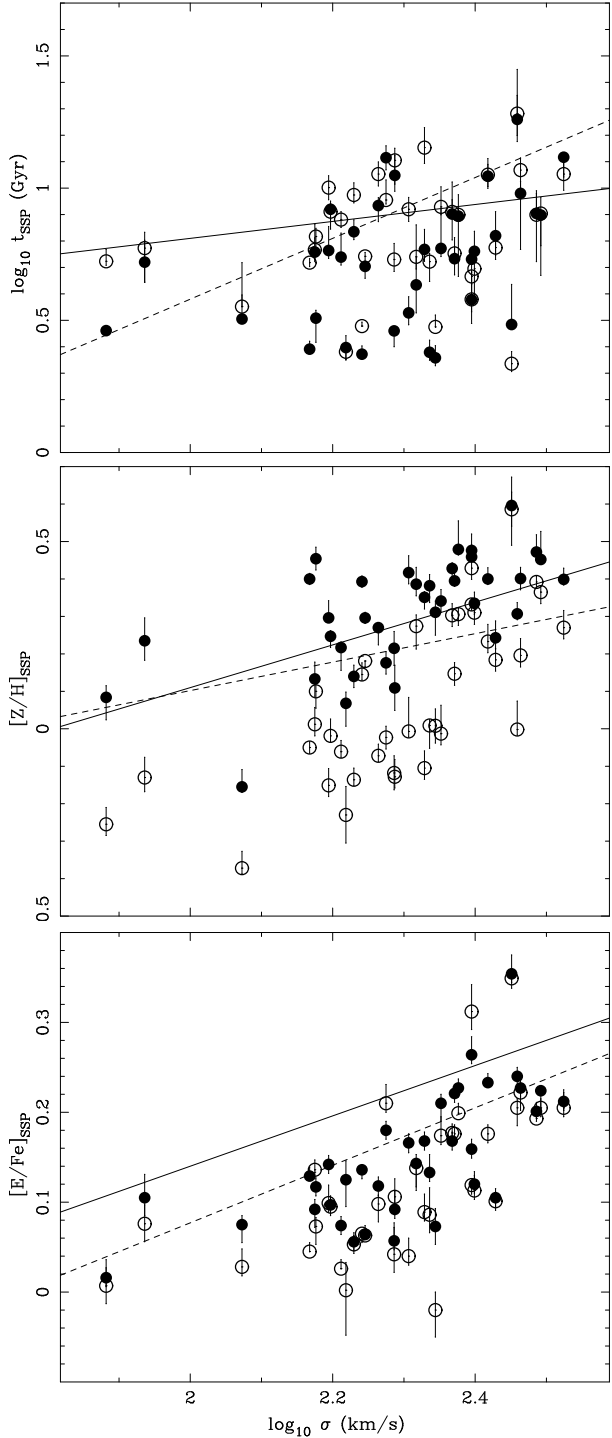}
\caption{\agessp, \metalssp\ and \abundssp\ vs. $\sigma$ within $R_e$/16 (filled circles) and $R_e$/2 (open circles). Solid and dashed lines represent Thomas et al.\ (2005) and Bernardi et al.\ (2006) fits respectively (see text).}
\label{stellarpop}
\end{figure}

The gas is typically extended. Fig.\ref{gasext} shows the H$\alpha$ flux detected between $R_e$/16 and $R_e$/2 -- i.e., $\Delta F_{H\alpha}=F_{H\alpha}(R_e/2)-F_{H\alpha}(R_e/16)$ -- normalised to the total $F_{H\alpha}$ and plotted against  the specific H$\alpha$ emission $f^*_{\rm H\alpha}(R_e/16)$. We define the latter as the H$\alpha$-to-stellar  flux ratio measured over 300 \AA\ centred on the emission line (values of $f^*_{\rm H\alpha}$ are given for both apertures in Table \ref{t3}). There are two (low-$f^*_{\rm H\alpha}$) galaxies where no emission is detected outside $R_e/16$ ($\Delta F_{H\alpha}=0$; NGC 3640 and NGC 7626), and five cases where all the emission comes from outside it ($\Delta F_{H\alpha}=F_{H\alpha}$; NGC 1439, NGC 2300, NGC 3610, NGC 7332 and NGC 7619). Generally, emission outside $R_e$/16 is the main contributor to the total. There also is a tendency of $H\alpha$-bright galaxies to have more concentrated emission.

In most of the spectra where H$\alpha$ is detected we also detect H$\beta$, \OIII\ and \NII. This allows us to study the ionisation state of the gas by using one of the flux-ratio diagnostic diagrams introduced by Baldwin, Phillips and Terlevich (1981). Fig.\ref{bpt} shows the ratio \OIII/H$\beta$ plotted versus \NII/H$\alpha$ for the $R_e$/16 (filled circles) and $R_e$/2 (open circles) apertures. This diagram allows us to separate different ionisation mechanisms as schematically indicated following Kauffmann et al.\ (2003). All galaxies are, over any aperture, LINER-like objects. We plot at \OIII/H$\beta$=0.1 galaxies where either \OIII\ or H$\beta$ are not detected but \NII and H$\alpha$ are. Also in these low-emission cases the ratio \NII/H$\alpha$ is outside the star-forming range.

Finally, Fig.\ref{hahi}, shows $f^*_{\rm H\alpha}(R_e/16)$ plotted against $M$(\HI) (in this case, filled and open circles represent \HI\ detections and upper limits respectively). At low \HI\ masses $f^*_{\rm H\alpha}$ is low while at high $M$(\HI) there is a wide range of ionised-gas content. The plot shows that a \HI\ mass larger than 10$^8$ \Msun\ is necessary (but not sufficient) for a galaxy to host bright ionised-gas. The same result is obtained using different emission lines or the $R_e$/2 aperture.

\subsection{Stellar populations}

Fig.\ref{stellarpop} plots the SSP-equivalent stellar parameters against central velocity dispersion for the $R_e$/16 (filled circles) and $R_e$/2 (open circles) aperture for all galaxies in the sample. Solid and dashed lines represent the linear fit of Thomas et al.\ (2005) and Bernardi et al.\ (2006) respectively, and are just meant as a comparison to previous results. In particular, we take the field, old galaxy population fit of Thomas et al., and the field, low redshift fit of Bernardi et al.

A few features stand out from Fig.\ref{stellarpop}. Firstly, there is a large scatter in all the plots and over both apertures. We find objects as young as 3 Gyr and as old as the Universe, with smaller systems being on average younger than bigger ones. \metalssp\ varies from half to three times solar and clearly increases when moving towards galaxies' inner region (as known since Peletier et al.\ 1990) as well as with $\sigma$. \abundssp\ ranges from solar to two times solar and also increases with $\sigma$. The slopes of the \metalssp-$\sigma$ and \abundssp-$\sigma$ relations are broadly consistent  with previous results. Both fits showed in the figure match well our metallicity  measurements within $R_e$/16 while our \abundssp\ measurements are systematically lower than what predicted by Thomas et al.\ fit. This is due to the use of different stellar population models. Clearly, our data imply a strong relation between \metalssp\ and \abundssp\ with metal-richer (and more massive) galaxies being also more $\alpha$/Fe-enhanced and therefore characterised by a shorter duration of the main star formation episode, as suggested by Thomas et al.\ (1999, 2005).

\begin{figure*}
\begin{center}
\includegraphics[width=17cm]{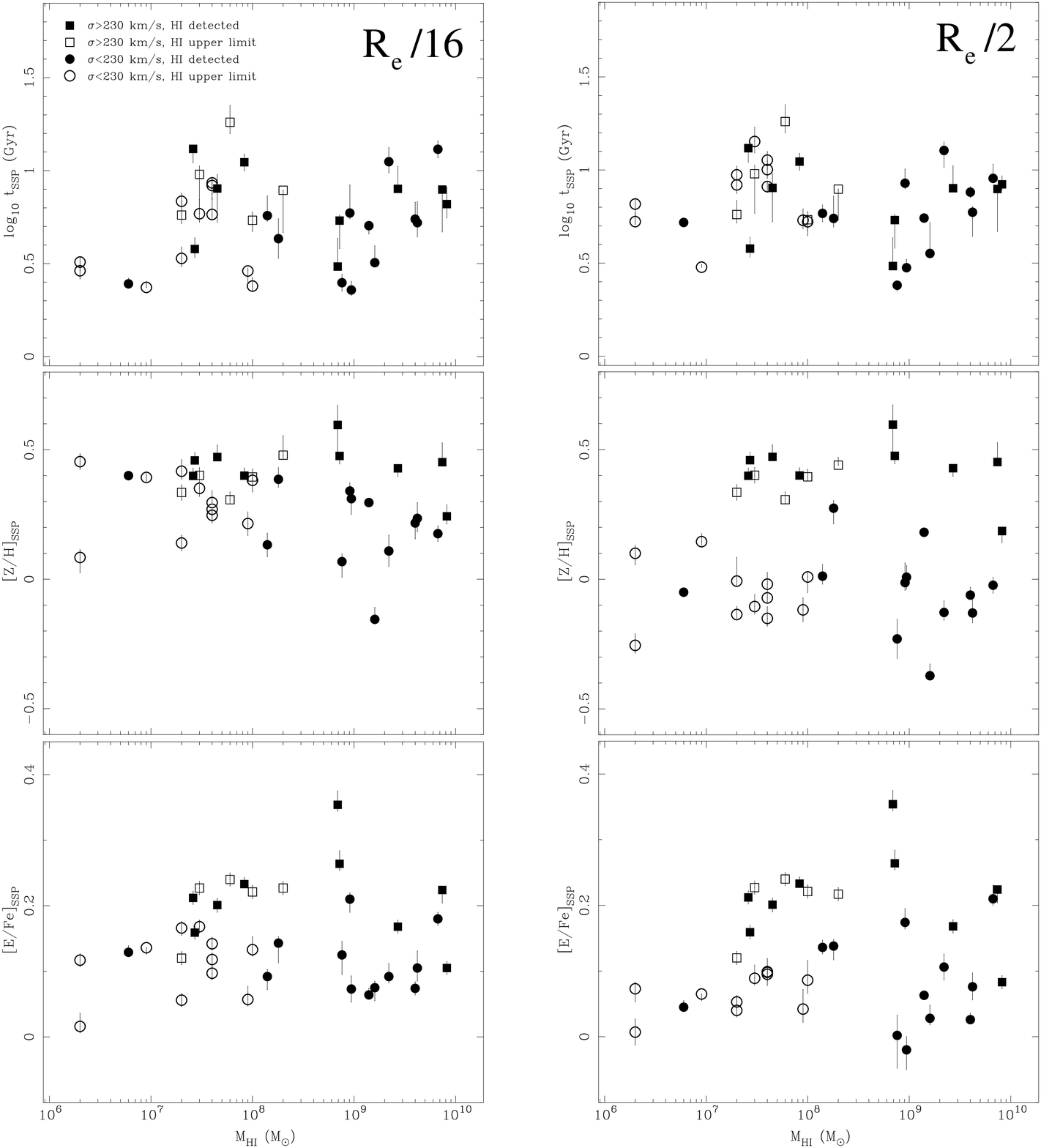}
\end{center}
\caption{\agessp, \metalssp\ and \abundssp\ in the inner $R_e$/16 (left) and $R_e$/2 (right) plotted against $M$(\HI). Filled and open symbols are \HI-detected and non-detected galaxies respectively. For the latter, the upper limit on $M$(\HI) is plotted. Circles and squares represent respectively low- and high-$\sigma$ galaxies as specified in the top-left plot.}
\label{sspRe16}
\end{figure*}

\subsection{Stellar populations and \HI}

Fig.\ref{sspRe16} shows \agessp, \metalssp\ and \abundssp\ plotted against $M$(\HI) for the $R_e$/16 and $R_e$/2 apertures. In both figures filled symbols represent \HI\ detections and open symbols \HI\ upper limits. Circles and squares mark galaxies with $\sigma\leq$230 km/s (low $\sigma$) and $\sigma\geq$230 km/s (high $\sigma$) respectively. The reason for a division in velocity dispersion will become clear below. The figure shows that the scatter visible in Fig.\ref{stellarpop} is present at any $M$(\HI), and that \agessp, \metalssp\ and \abundssp\ do not depend on the \HI\ mass. On the other hand, we do find interesting trends in the spatial distribution of stellar populations.

\begin{figure}
\includegraphics[width=8cm]{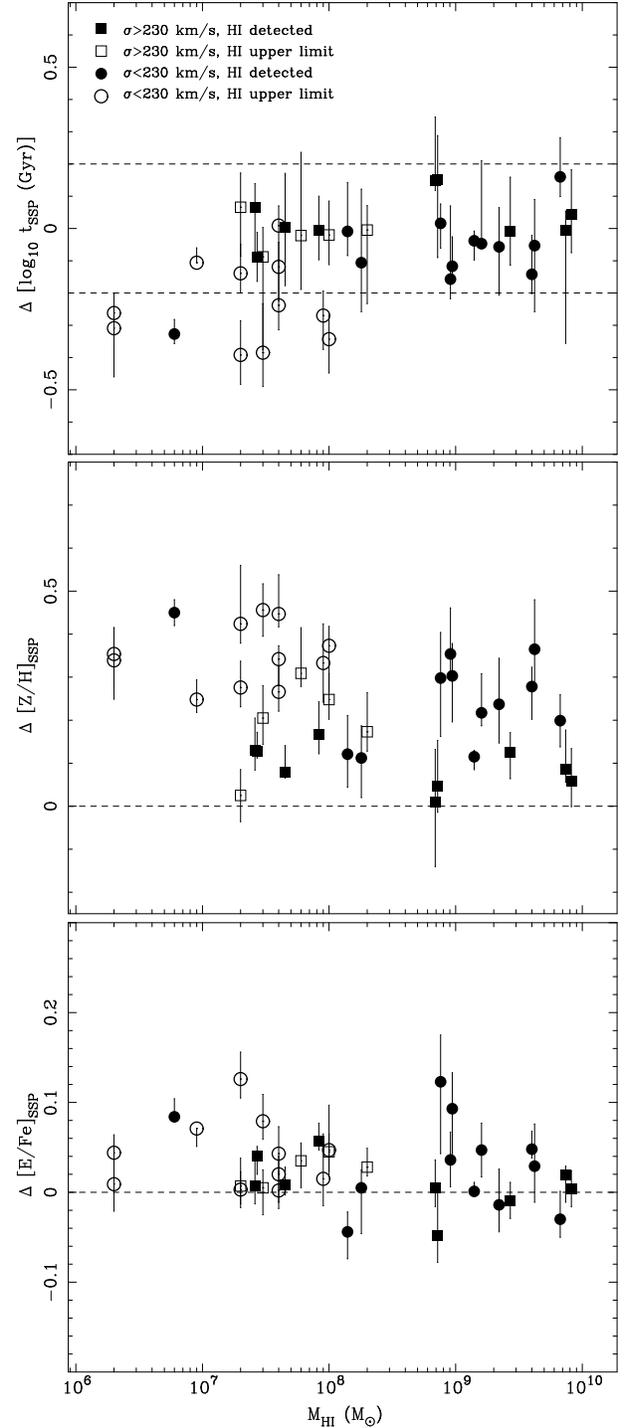}
\caption{Variation of \agessp, \metalssp\ and \abundssp\ when going from the $R_e$/2 aperture to the $R_e$/16 one. Markers are as in Fig.\ref{sspRe16}. For each parameter, a negative $\Delta$ implies a decrease in that parameter when reducing the aperture radius from $R_e$/2 to $R_e$/16.}
\label{deltassp}
\end{figure}

Fig.\ref{deltassp} shows the variation of all SSP-equivalent parameters when moving from the inner $R_e$/2 to the inner $R_e$/16 plotted against $M$(\HI). Graphic markers are the same as in Fig. \ref{sspRe16}. We find that 8 out of 20 early-type galaxies with $M$(\HI)$\leq10^8$\Msun\ exhibit a central drop in age ($\Delta log_{10}$\agessp $\leq -0.2$; these are NGC 1439, NGC 2549, NGC 3193, NGC 3640, NGC 7332, NGC 7457, NGC 7585 and NGC 7600) while none of the 17 objects with $M$(\HI)$\geq10^8$\Msun\ do. In particular, such $\Delta t$-$M$(\HI) connection does not hold for high-$\sigma$ objects, none of which are centrally rejuvenated independently on the neutral hydrogen mass. On the contrary, at low $\sigma$, 2/3 of the \HI-poor galaxies exhibit a central age drop (this fraction drops to $\sim$55\% if we use $\sigma$=250 km/s as a dividing line) while none of the  \HI-rich object do.

\begin{figure}
\includegraphics[angle=270,width=8cm]{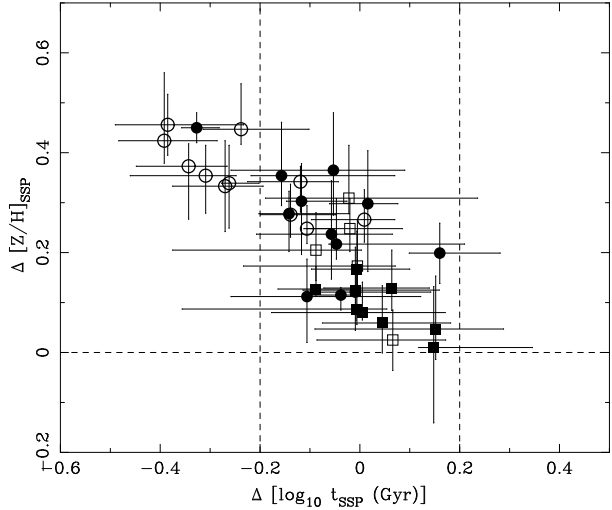}
\caption{\metalssp\ variation plotted against \agessp\ variation when going from $R_e$/2 aperture to the $R_e$/16 one. Markers are as in Fig.\ref{sspRe16}.}
\label{deltadelta}
\end{figure}

As anticipated in the previous section, we also observe a clear increase in \metalssp\ when moving towards the centre of galaxies, while \abundssp\ does not change much. Furthermore, by comparing $\Delta$\agessp\ and $\Delta$\metalssp, Fig.\ref{deltadelta} shows that galaxies with a central age drop have larger central \metalssp\ increase compared to the other objects.

\subsection{H$\beta$ correction and the case of IC 4200}

Our analysis relies heavily on a correct determination of H$\beta$ absorption. As many galaxies in the sample are contaminated by ionised-gas emission lines, including H$\beta$, we check to what extent our result is driven by the line infill correction obtained with GANDALF. We list the H$\beta$ index corrections for all galaxies in Table \ref{t3}.

Fig.\ref{hahi} already suggests that rejuvenated objects, being \HI-poor, are also poor in ionised gas. Their stellar population parameters must therefore be little affected by gas-subtraction errors. Indeed, only two of the galaxies with $\Delta log_{10}$\agessp $\leq -0.2$ have non-zero H$\beta$ corrections. These are NGC 3193 and NGC 7585. For the first, the corrections are 0.24 and 0.14 \AA\ along the $R_e$/16 and $R_e$/2 apertures respectively. For the second the correction is 0.08 \AA\ along the $R_e$/16 aperture and null along the $R_e$/2 one. Concerning the 29 galaxies with $|\Delta log_{10}$\agessp $|\leq 0.2$, the H$\beta$ correction is below 0.5 \AA\ in all but 8 and 6 galaxies over the $R_e$/16 and $R_e$/2 aperture respectively, and zero in half of the cases.

Stellar population models show that an error of 0.1 \AA\ in the H$\beta$ correction, which would represent a large fraction of the infill correction in most of our galaxies, translates into an error of 0.02, 0.05 and 0.1 dex in $log_{10}$\agessp\ at 1, 5 and 10 Gyr respectively (this can be understood from Fig.\ref{grid}, top panel). We therefore conclude that our result, and in particular the fact that many low-$\sigma$, \HI-poor galaxies are centrally rejuvenated, is not significantly affected by the H$\beta$ correction.

The importance of a correct subtraction of the ionised-gas contamination is well illustrated by the case of IC 4200. This is among the few galaxies with substantial H$\beta$ correction and was studied in details in Serra et al.\ (2006; S06). In that work we modelled the ionised-gas emission in a different, simpler way. The main differences with respect to the present study are that in S06 we found the best-fitting BC03 SSP model instead of the best \emph{combination} of MILES SSP models; and that we computed the best Gaussian fit to the residual emission lines successively rather than simultaneously (see S06 for details). The different method resulted in a much younger age than what reported here (see Table \ref{t5}). The reason is that the best-fitting stellar model obtained now is older  (giving also an improved fit). An additional effect is that in S06, unlike now, we could not properly subtract the \NI\ emission. When this is done, the red continuum of Mg$b$ decreases so that the index itself becomes weaker, causing an additional age increase (this can be understood from Fig.\ref{grid}). We consider the new result more reliable.

\section{Discussion}

\subsection{Ionised gas}

We find two very clear results with respect to the ionised gas. The first is that the gas is, with few exceptions, extended and characterised by LINER-like line ratios at any distance from the centre of the host galaxy (H$\alpha$ and \NII\ lines are crucial to reach this conclusion). Although our sample is not representative, this result is consistent with previous studies on the spatial ionised-gas distribution in early-type galaxies (Sarzi et al.\ 2006), and on the incidence of LINER-like systems in large samples of emission-line E/S0's (3 out of 5 according to Yan et al.\ 2006).

As explained in the review by Filippenko (2003), LINER-like line ratios can be produced by different mechanisms: a central AGN; heating via shocks; ionisation by old stars; cooling flow of hot gas. Therefore, although LINERs were first introduced as a class of systems with nuclear emission lines, it is in principle possible to explain the observed diffuse LINER-like emission. In our case, the most important indication is that the ionisation is $not$ caused simply by low-level residual star formation.

The second finding is that only very \HI-rich galaxies ($M$(\HI)$\geq10^8$\Msun) can host strong ionised-gas emission. This is not a trivial result. The \HI\ is typically distributed out to very large radii (many tens of kpc; see Morganti et al.\ 2006; Oosterloo et al.\ 2007), while the ionised gas is being observed in the inner few kpc of the stellar body. Therefore, the two gas phases are not necessarily physically linked; and indeed there are examples of galaxies with large amounts of \HI\ but no or little ionised gas (see Fig.\ref{hahi}). The fact that only \HI-rich galaxies show strong emission could imply that, at least in some cases, the ionisation mechanism is provided by the shocks experienced by gas flowing from the large scale \HI\ systems towards the inner stellar body. Such a connection would be consistent with the finding of Morganti et al.\ (2006) that ionised- and \HI-gas velocity fields are contiguous in systems with regular neutral-hydrogen rotation. In this sense, it will be interesting to compare in more detail the kinematics of \HI\ and ionised gas in our sample. This will be the subject of future work.

\subsection{Stellar populations}

Concerning the stellar populations, we confirm the general increase of \agessp, \metalssp\ and \abundssp\ with $\sigma$, with a large scatter especially in the \agessp-$\sigma$ relation. On the other hand, we do not find \emph{any} trend in the stellar age and chemical composition with $M$(\HI). \HI-rich galaxies are not systematically younger/older than \HI-poor ones, nor there is a difference in metallicity or abundance ratio. At any $M$(\HI), all stellar population parameters are characterised by a large scatter. However, we do find that, in a statistical sense and over the present sample, and with the important exception of the most massive galaxies ($\sigma\geq$230 km/s), \HI-poor galaxies show a central age-drop while \HI-rich galaxies do not; and that the drop in age is associated to a larger-than-average central metallicity increase.

To interpret this result it is important to realise that the analysis technique used in the present study gives only one value for the stellar age and chemical composition in a galaxy, as if all stars were formed at the same time and from gas clouds with the same metallicity. However, stars within a galaxy almost certainly have a spread in age and in the abundance of individual elements. In such a more realistic situation \metalssp\ and \abundssp\ track the $V$-band luminosity-weighted chemical composition while \agessp\ is always strongly biased towards the age of the youngest stars (see Serra \& Trager 2007 for a thorough discussion of these systematics). For example, a 5/100 mass ratio between a 1-Gyr-old and a 10-Gyr-old population can drive \agessp\ down to 2.5 Gyr (at solar [Z/H] and [E/Fe]). This bias fades very quickly as the young stellar component ages: 1 Gyr later, the same system would have \agessp=7 Gyr; 2 Gyr later, \agessp=10 Gyr.

It is therefore likely that the central age drop in low-$\sigma$ \HI-poor objects is indeed caused by a central stellar sub-component younger than the underlying population. Furthermore, because the age-drop is associated to a larger increase in \metalssp\ than in the rest of the sample, we conclude that the young stellar population in the centre was formed from enriched material.

Given the size of the sample and the magnitude of the effect described, this result needs being strengthened by further investigation. In particular, we stressed in the introduction that gas plays an important role in shaping the orbital structure of galaxies. Therefore, the analysis of the stellar and gas kinematics of galaxies in the sample will be an important complement to the work presented here. Such analysis is currently being performed and is the subject of a forthcoming paper. For the moment, it is worthwhile trying to understand how the picture described above could arise.

The young stellar population in the centre of \HI-poor objects is likely to have formed in situ (the less convincing alternative would be that, unlike \HI-rich galaxies, a large fraction of \HI-poor E/S0's accreted a young and metal-rich stellar system onto their very inner regions). Therefore, having hosted central star formation in their recent past, more than half of the low-$\sigma$ \HI-poor systems must have contained a significant gas mass from which to form stars; i.e., \emph{they must have been \HI-rich themselves}. Clearly, the gas from which stars formed did not necessarily have the same properties (morphology, kinematics, density) as that observed around currently-\HI-rich systems. Yet, it must have been there at some point. And most importantly, in order to justify a significant drop in the central age of \HI-poor systems, it must have been as massive as some percent of the total galaxy stellar mass. This is of the same order of magnitude of the \HI\ mass detected around \HI-rich galaxies (10$^8$ to $\sim$10$^{10}$ \Msun).

\HI-rich and  more than half of the \HI-poor low-$\sigma$ systems might have therefore contained similar gas masses in the past. What caused the now-\HI-poor galaxies to use their gas supply to form highly concentrated stars, and the now-\HI-rich objects not to do so?

\subsection{Mergers of galaxies}

According to the picture for the formation of early-type galaxies sketched in the introduction, mergers are a necessary ingredient to transform disk galaxies into spheroids from z$\sim$1 to now, and account for the E/S0 luminosity-function evolution at magnitudes fainter than $M_B\leq-21$. With respect to the results just described, it is interesting to note that mergers of disk galaxies can result in gas-rich or gas-poor remnants depending on the details of the merging process. In particular, it is important whether or not the gas of the progenitor galaxies can retain its angular momentum with respect to the remnant (e.g., Di Matteo et al.\ 2007).

If most of the gas loses its angular momentum (like in a retrograde encounter, in which the spins of the two disk-galaxies lie along opposite directions) it falls towards the very centre of the remnant, where it is consumed in an efficient burst of star formation. In this case no/little gas is left at large radii so the remnant is \HI-poor. Furthermore, because of the central star formation, the average stellar age should drop in the inner kpc.

On the other hand, if gas conserves its angular momentum (for example in a prograde encounter, in which the two spins are aligned) large tidal-tails form and up to 60\% of the progenitors' gas is spread at large radii (Barnes 2002). This can later be re-accreted and form massive, extended \HI\ structures like the ones observed around \HI-rich galaxies. Simulations show that also in prograde-like mergers gas infall and central star formation occur. However, given a same progenitor pair, the gas available for star formation within the stellar body is less than for a retrograde-like encounter. Furthermore, gas in a prograde encounter has generally higher angular momentum than in a retrograde one so that it (and any related star formation) is less concentrated towards the centre.

Fig. 34 in Di Matteo et al.\ (2007) shows this effect quite clearly by plotting the Lagrangian radii of the gas+new stars \footnote{I.e., the radii containing a given fraction of the total gas+new stars mass.} for both a retrograde and a prograde merger. In the former case, all Lagrangian radii are smaller at coalescence than the ones of the isolated progenitors, with the 50\% and 75\% radii being reduced by 1-2 orders of magnitude. On the contrary in a prograde merger the 10\%, 25\% and 75\% Lagrangian radii are larger at coalescence than for an isolated progenitor, while only the 50\% radius is significantly smaller, as during a retrograde encounter. Therefore, it is reasonable to presume that \HI-rich remnants have a smaller (if any) age drop in their very centre.

Indeed, if disk mergers play a significant role in early-type galaxy formation (as argued by Bell et al.\ 2004b and Faber et al.\ 2007) one might expect to see what Fig.\ref{deltassp} suggests: \HI-poor objects (or at least a fraction of them) with a strong central age drop and \HI-rich systems without it. Furthermore, since most of the red-sequence luminosity-function evolution occurs at magnitudes fainter than $M_B$=$-21$ (Brown et al.\ 2007), this effect should not involve massive galaxies. Interestingly, Brown et al.\ approximate magnitude limit corresponds to a velocity dispersion of $\sim$250 km/s, above which we indeed do not find galaxies with a central age drop associated with a low \HI\ mass.

Clearly, even if the described mechanism plays an important role in E/S0 formation, some scatter is expected. Firstly, any galaxy that did not form via a gas-rich merger would not fall in this picture. Furthermore, a large number of initial conditions plays a role in determining the final result of a merger (e.g., progenitors' gas content, orbital parameters, spin alignment, mass ratio, progenitors' bulge/disk ratio). The environment is also an important factor; at higher density, not probed by our sample, the IGM affects galaxies gas phase in a variety of ways (e.g., gas stripping and heating), so that the picture in a cluster of galaxies is likely to be quite different. Finally, as mentioned above, the longer the time passed since the merger and thus since the burst of star formation, the weaker will be the impact of the young stellar component on the SSP-equivalent parameters. With respect to this, we must keep in mind that many \HI-rich systems are characterised by settled neutral-hydrogen configurations (e.g., Morganti et al.\ 2006; Serra et al.\ 2006; Oosterloo et al.\ 2007), indicating that the merger (if any) occurred many Gyr ago.

We note that this interpretation implies the existence of objects formed via mergers among both \HI-rich and \HI-poor galaxies. This is reasonable as at any $M($\HI) we find young systems, i.e., systems that are likely to have hosted recent star formation. At any \HI-mass star formation could have  been triggered by a gas-rich merger, \emph{with the large-scale \HI\ survival being linked to the spatial distribution of the star formation during the merging process}.

As a word of caution, we must point out that such interpretation is based on a comparison of galaxy stellar populations to simulations of gas-rich mergers. Unfortunately, while the dynamics of both gas and stars during a merger might be well described by simulations, a full understanding of the physics of star formation during these processes is still lacking. For example, there are suggestions that star formation during galaxy interactions might be induced by shocks instead of being simply dependent on the gas density (Barnes 2004). As a consequence, the discussion above has to be rather qualitative. Theoretical improvement will be crucial in enabling a more accurate, quantitative approach.

Finally, the fact that the young population is metal rich is consistent with it having formed from gas that belonged to the interstellar medium of a disk galaxy; i.e., already enriched material. As we argue below, this is a strong argument against accretion of gas from the intergalactic medium being responsible for the existence of \HI-rich E and S0's.

\subsection{Structural properties and IGM accretion}

The above interpretation has the advantage of being linked to a sound, observationally-based picture for the evolution of early-type galaxies since z$\sim$1. However, it is important to understand whether it is the only possible one.

Fig.\ref{deltassp} can be interpreted  under the general scheme that if gas (however accreted) cannot retain its angular momentum it falls towards the centre of a galaxy, leaving it devoid of large-scale gaseous structures. There it may trigger a burst of star formation which in turn causes a stellar-age drop with respect to the outer regions. This could be caused by any structural property that allows gas angular-momentum removal, including non-axisymmetric mass distributions. The gas fate would then be determined by some intrinsic galaxy properties rather than the way the gas was accreted (e.g., prograde- vs. retrograde-like galaxy encounters).

Among non-axisymmetric structures, triaxial bulges or halos are not straightforward to identify. In any case their triaxiality is never extremely strong, and halos in particular are triaxial on a large scale compared to disks so that they are probably seen as nearly spherical. At the same time, strong spiral arms are not present in this sample as our galaxies are all early-type. Bars on the other hand could be hosted by our targets. It is possible that centrally-rejuvenated \HI-poor systems contain a bar that caused gas inflow and, possibly, successive gas depletion via star formation. On the contrary, (strong) bars should not be found in \HI-rich systems.

There are a few problems with this picture. Firstly, only two of the galaxies in the sample are reported by previous observers to contain a bar: NGC 5018 (Corwin et al.\ 1985) and NGC 7332 (Falc\'{o}n-Barroso et al.\ 2004). The former is actually a uniformly young (2-3 Gyr old) \HI-rich object while the latter has a very modest amount of \HI\ and indeed a younger central stellar population. NGC 7332 is therefore the only case in which a bar might be the cause of central rejuvenation. Three more systems are suspected of hosting a bar (NGC 0596 and NGC 0636, Nieto et al.\ 1992; NGC 7457, Michard \& Marchal 1994) but no strong evidence is given. Furthermore, of these, only NGC 7457 hosts a young central stellar component. Finally, none of the galaxies shows obvious signs of a bar in 2MASS images. This means that only weakly-barred objects, if barred at all, are present in the sample \footnote{A possible complication comes from the fact that bars may cyclically form and be destroyed in galaxies, so that a lack of bar detection does not completely rule out its past existence (see Combes et al.\ 2007).}.

Secondly, along with gas inflow bars also re-distribute material around the location of the outer Linblad resonance, so that a significant fraction of the initial available gas should still be found at large radii (P\'{e}rez \& Freeman 2006). This problem is likely to be even more relevant if the gas is initially spread over very large distances as observed in the \HI-rich E/S0's. Therefore, a link between central rejuvenation and lack of large-scale \HI\ is not an obvious outcome of a central bar.

Furthermore, it is not at all clear that a bar would be capable of transporting many times 10$^9$ \Msun\ of gas towards the galaxy centre, making a \HI-rich object become \HI-poor. This might be the case for very strong bars, but the existence of barred disk galaxies with large disks of rotating \HI\ already suggests that efficient gas removal is not the rule.

Finally, even if bars were responsible of the gas inflow, it would remain to be explained how the gas was initially accreted by both \HI-rich and centrally-rejuvenated \HI-poor galaxies. Accretion of gas from the intergalactic medium (IGM) is a viable way of accumulating large masses of gas around galaxies. In particular, simulations show that a fraction of the gas is normally accreted via a cold mode, i.e. it is not shock-heated to the virial temperature of the accreting halo but stays below 10$^5$ K (Binney 1977; Fardal et al.\ 2001; Keres et al.\ 2005, Birnboim et al.\ 2007). Depending on halo mass and environment, as much as 10$^{10}$ \Msun\ of gas can be accreted via the cold mode and, if the environment allows it, cool to the neutral state in a reasonable time scale ($\sim$1 Gyr provided a small enrichment from stellar processes within the galaxy; see the discussion in Serra et al.\ 2006).

Cold accretion can therefore explain the flow of substantial amounts of gas onto early-type galaxies. However, the young stellar population in the centre of many \HI-poor E/S0's is hard to interpret in terms of bar-induced inflow of IGM material. The reason is that, even if slightly enriched by stellar processes within the accreting galaxy, this gas cannot be as metal-rich as our stellar-population analysis suggests (many times solar). Therefore, it is possible that some \HI-rich systems have accreted their gas from the IGM during a prolonged, dynamically-quiet phase. But it seems unlikely that this process has played the dominant role in bringing gas around E/S0's.

\section{Conclusions}

We have presented a study of the stellar population and ionised gas content of a sample of 39 field early-type galaxies as a function of their \HI\ mass. We find that 60\% of them host diffuse, LINER-like, ionised gas emission. \HI-poor galaxies tend to contain little ionsed gas, while \HI-rich systems can have high specific emission as well as no emission at all. A large \HI\ mass appears therefore to be a necessary (but not sufficient) condition to host large warm gas reservoirs.

The stellar populations, analysed by means of line-strength indices, show increasing age, metallicity and $\alpha$/Fe with $\sigma$ but with a large scatter (in particular the age). None of these parameters depends on the \HI\ mass. With respect to their variation across galaxies' stellar body, we find a clear increase in metallicity when moving towards the centre. 

The most interesting result is that in galaxies with velocity dispersion below 230 km/s a low neutral-hydrogen content is associated, in 2/3 of the cases, to a central stellar-age drop by a factor of at least 1.7. We interpret such age variation in terms of a central stellar component younger than the underlying one. \HI-rich objects never show such age gradients (although they can be uniformly young). On the other hand, in more massive galaxies no $\Delta t$-$M$(\HI) relation is found.

We find that this result can be interpreted within the picture of the evolution of E/S0's that has emerged from recent high-redshift surveys. These require that a large fraction of $M_B\leq-21$-spheroids are formed via disk-galaxy mergers from z$\sim$1 to now. It is possible that during a merger the large-scale gas survival is inversely proportional to the gas-infall efficiency and therefore to the related star-formation intensity and concentration. This would explain why a large fraction of low-$\sigma$ galaxies without large, massive \HI\ distributions host a young central stellar component while \HI-rich galaxies have more uniform (even when young) populations.

Given the size of our sample and the magnitude of the described $\Delta t$-$M$(\HI) relation, this result needs further investigation. Three aspects need particular attention. Firstly, it is crucial to achieve a quantitative, theoretical understanding of the fate of gas during mergers, and in particular of the intensity and spatial distribution of any star formation that it might host. Secondly, the different fate of gas possibly revealed by the stellar population distribution across galaxies might at the same time leave signatures in their stellar kinematics; we are currently analysing the data presented here to explore this possibility. Finally, a large, representative sample of early-type galaxies should be observed at both 21 cm and optical wavelengths in order to perform the same kind of analysis presented here. This would not only enable to confirm and possibly strengthen our result, but also to understand what fraction of E/S0's show sign of recent merging assembly and therefore how important gas-rich mergers are for the formation of local early-type galaxies.

\begin{acknowledgements}

The authors would like to thank Thijs van der Hulst and Jacqueline van Gorkom for interesting discussions; Guy Worthey for providing index responses to abundance-ratio variations in advance of publication; Marc Sarzi for support in GANDALF usage; and Patricia S\'{a}nchez-Bl\'{a}zquez for providing the transformation from the MILES to the Lick/IDS system. PS wishes to thank Isabel P\'{e}rez for stimulating conversations. This research is based on data obtained with the William Herschel Telescope, which is operated on the island of La Palma by the Isaac Newton Group in the Spanish Observatorio del Roque de los Muchachos of the Instituto de Astrof\'{i}sica de Canarias; and with the New Technology Telescope operated by the European Southern Observatory at La Silla, Chile. This research made use of GANDALF, which is developed by the SAURON team and is available from the SAURON website (www.strw.leidenuniv.nl/sauron). We also aknowledge the usage of the HyperLeda database (http://leda.univ-lyon1.fr); and of NASA/IPAC Extragalactic Database (NED) which is operated by the Jet Propulsion Laboratory, California Institute of Technology, under contract with the National Aeronautics and Space Administration.

\end{acknowledgements}

\begin{table*}
\begin{center}
\caption{Ionised-gas emission-lines flux $F$ (arbitrary units), H$\beta$-index corrections $\Delta$H$\beta$, and specific H$\alpha$ emission $f^*_{\rm H\alpha}$.}
\label{t3}
\begin{tabular}{lrrrrrrrrrrrrrr}
\hline
\hline
\noalign{\smallskip}
\multicolumn{1}{c}{galaxy}      & \multicolumn{2}{c}{$F$(H$\beta$)} & \multicolumn{2}{c}{$F$(\OIII)} & \multicolumn{2}{c}{$F$(H$\alpha$)} & \multicolumn{2}{c}{$F$(\NII)} & & \multicolumn{2}{c}{$\Delta$H$\beta$ (\AA)} & & \multicolumn{2}{c}{$f^*_{\rm H\alpha}$} \\
            & $R_e$/16 & $R_e$/2           & $R_e$/16 & $R_e$/2        & $R_e$/16 & $R_e$/2         & $R_e$/16 & $R_e$/2         & & $R_e$/16 & $R_e$/2         & & $R_e$/16 & $R_e$/2 \\ 
\noalign{\smallskip} \hline \noalign{\smallskip}
\noalign{\smallskip} \noalign{\smallskip}
\multicolumn{15}{c}{\emph{NTT/EMMI spectroscopy}}\\
      ESO 092-21  &    0.1  &    2.0  &    0.2  &      3.0  &   0.2  &      7.6  &      0.2  &      5.6 & & 0.47 & 0.44 & &  3.2 & 3.0 \\
      ESO 140-31  &    0.1  &    0.9  &    0.3  &      3.0  &   0.4  &      3.7  &      0.5  &      4.0 & & 0.54 & 0.36 & &  3.1 & 2.7 \\
      ESO 381-47  &     -   &    1.4  &    0.4  &      2.4  &   0.4  &      3.1  &      0.7  &      2.7 & & -    & 0.36 & &  0.7 & 1.6 \\
       IC 4200    &    0.2  &    3.1  &    0.4  &      4.5  &   1.5  &     15.8  &      2.2  &     19.0 & & 1.51 & 1.26 & & 10.6 & 8.4 \\
       IC 4889    &    0.1  &    2.4  &    0.3  &      5.9  &   0.6  &     10.3  &      0.9  &     13.8 & & 0.33 & 0.31 & &  2.9 & 2.5 \\
      NGC 1490    &     -   &     -   &     -   &       -   &    -   &       -   &       -   &       -  & & -    & -    & &  -   & -   \\
      NGC 1947    &    1.6  &    9.1  &    2.7  &     13.8  &   9.0  &     38.0  &      8.2  &     30.1 & & 0.99 & 0.70 & &  8.6 & 4.9 \\
      NGC 2434    &     -   &     -   &    0.9  &      4.5  &   1.2  &      5.7  &      1.9  &      5.6 & & -    & -    & &  1.0 & 0.8 \\
      NGC 2904    &     -   &     -   &     -   &       -   &    -   &       -   &       -   &       -  & & -    & -    & &  -   & -   \\
      NGC 3108    &    1.2  &    5.5  &    3.7  &     15.1  &   5.1  &     20.5  &      8.0  &     23.0 & & 0.46 & 0.40 & &  3.0 & 2.9 \\
\noalign{\smallskip} \noalign{\smallskip}
\multicolumn{15}{c}{\emph{WHT/ISIS spectroscopy}}\\
      NGC 0596    &     -   &     -   &     -   &     14.3  &    -   &       -   &     14.3  &     36.2 & & -    & -    & &  -   & -   \\
      NGC 0636    &     -   &     -   &     -   &       -   &    -   &       -   &     34.8  &       -  & & -    & -    & &  -   & -   \\
      NGC 1426    &     -   &     -   &     -   &       -   &    -   &       -   &       -   &     60.0 & & -    & -    & &  -   & -   \\
      NGC 1439    &     -   &     -   &     -   &       -   &    -   &     14.4  &      7.2  &     21.1 & & -    & -    & &  -   & 0.4 \\
      NGC 2300    &     -   &     -   &     -   &       -   &    -   &     23.8  &       -   &       -  & & -    & -    & &  -   & 0.4 \\
      NGC 2534    &    3.5  &   11.9  &    3.7  &      9.5  &   5.2  &     73.5  &     15.6  &     42.4 & & 1.19 & 1.07 & & 10.9 & 8.8 \\
      NGC 2549    &     -   &     -   &    3.0  &     18.6  &    -   &       -   &     10.9  &     40.4 & & -    & -    & &  -   & -   \\
      NGC 2768    &    6.3  &   25.2  &   11.9  &     57.5  &   2.1  &    152.2  &     65.9  &    238.2 & & 0.59 & 0.33 & &  4.8 & 2.5 \\
      NGC 2810    &   10.7  &   31.5  &   13.3  &     29.9  &   8.7  &    131.4  &     79.7  &    158.4 & & 1.79 & 1.25 & & 12.0 & 6.8 \\
      NGC 3193    &    3.8  &   16.8  &    6.9  &     33.9  &   2.0  &     56.9  &     16.8  &     68.4 & & 0.24 & 0.14 & &  1.2 & 0.8 \\
      NGC 3610    &     -   &     -   &    1.3  &       -   &    -   &     90.9  &     12.7  &     80.3 & & -    & -    & &  -   & 1.3 \\
      NGC 3640    &     -   &     -   &     -   &       -   &   5.0  &       -   &     23.1  &     68.0 & & -    & -    & &  0.4 & -   \\
      NGC 3998    &  130.1  &  229.3  &  126.2  &    178.2  &   8.5  &    548.2  &    675.8  &    953.0 & & -    & -    & & 14.4 & 6.2 \\
      NGC 4026    &     -   &   30.6  &    7.4  &     63.0  &   8.8  &     78.6  &     12.8  &     53.9 & & -    & 0.16 & &  0.4 & 0.5 \\
      NGC 4125    &    8.9  &   44.1  &   15.1  &     67.5  &   9.5  &    244.4  &     98.9  &    455.2 & & 0.47 & 0.33 & &  3.4 & 2.4 \\
      NGC 4278    &   71.7  &  165.6  &   94.8  &    211.4  &   6.7  &    463.8  &    356.8  &    616.7 & & 3.04 & 1.37 & & 15.2 & 5.2 \\
      NGC 4406    &     -   &     -   &     -   &       -   &    -   &       -   &       -   &       -  & & -    & -    & &  -   & -   \\
      NGC 4472    &     -   &     -   &     -   &       -   &    -   &       -   &       -   &       -  & & -    & -    & &  -   & -   \\
      NGC 5018    &     -   &     -   &    4.5  &     10.5  &   9.7  &     88.2  &     61.7  &    184.7 & & -    & -    & &  2.8 & 0.6 \\
      NGC 5173    &   11.0  &   22.1  &   18.5  &     31.3  &   5.8  &    126.9  &     35.5  &     75.1 & & 1.66 & 1.07 & & 13.2 & 8.5 \\
      NGC 5322    &     -   &     -   &     -   &       -   &    -   &       -   &     59.0  &    210.7 & & -    & -    & &  -   & -   \\
      NGC 5903    &    1.9  &   12.5  &    4.1  &     23.4  &   6.4  &     92.3  &     17.3  &     81.0 & & 0.25 & 0.32 & &  2.8 & 3.0 \\
      NGC 7052    &    5.7  &   15.8  &    3.9  &      0.8  &  30.0  &    115.6  &     61.8  &     68.1 & & -    & -    & &  5.2 & 2.0 \\
      NGC 7332    &     -   &     -   &    6.7  &     44.7  &    -   &     48.2  &     14.9  &     59.1 & & -    & -    & &  -   & 0.4 \\
      NGC 7457    &     -   &     -   &    0.8  &       -   &    -   &       -   &       -   &       -  & & -    & -    & &  -   & -   \\
      NGC 7585    &    0.7  &     -   &    2.3  &      6.9  &   8.3  &     49.7  &     12.8  &     63.5 & & 0.08 & -    & &  1.4 & 0.7 \\
      NGC 7600    &     -   &     -   &     -   &       -   &    -   &       -   &      9.6  &       -  & & -    & -    & &  -   & -   \\
      NGC 7619    &     -   &     -   &     -   &       -   &    -   &     42.7  &       -   &     61.2 & & -    & -    & &  -   & 0.5 \\
      NGC 7626    &     -   &     -   &     -   &       -   &   6.0  &       -   &     10.6  &     13.0 & & -    & -    & &  0.9 & -   \\
\noalign{\smallskip}
\hline
\end{tabular}
\end{center}
Flux of the four ionised-gas emission lines used in this paper for all galaxies and apertures (no flux value is given when the detection criterion is not satisfied; see Sec.\ref{gandalf}); correction on the H$\beta$ line-strength index determined by the detected H$\beta$ emission; and specific H$\alpha$ emission for each galaxy and aperture. For the definition of $f^*_{\rm H\alpha}$ see the Sec.\ref{fstardefinition}.
\end{table*}

\begin{table*}
\begin{center}
\caption{Lick/IDS line-strength indices}
\label{t4}
\begin{tabular}{lcccccccc}
\hline
\hline
\noalign{\smallskip}
\multicolumn{1}{c}{galaxy}      & \multicolumn{2}{c}{H$\beta$ (\AA)} & \multicolumn{2}{c}{Mg$b$ (\AA)} & \multicolumn{2}{c}{Fe5270 (\AA)} & \multicolumn{2}{c}{Fe5335 (\AA)} \\
            & $R_e$/16 & $R_e$/2           & $R_e$/16 & $R_e$/2        & $R_e$/16 & $R_e$/2         & $R_e$/16 & $R_e$/2 \\ 
\noalign{\smallskip} \hline \noalign{\smallskip}
\noalign{\smallskip} \noalign{\smallskip}
\multicolumn{9}{c}{\emph{NTT/EMMI spectroscopy}}\\
       ESO 092-21 &  1.86 $\pm 0.09$ &  2.05 $\pm 0.07$ &  4.19 $\pm 0.11$ &  3.31 $\pm 0.09$ &  3.00 $\pm 0.11$ &  2.41 $\pm 0.09$ &  2.87 $\pm 0.13$ &  2.48 $\pm 0.10$\\
      ESO 140-31 &  1.83 $\pm 0.06$ &  1.82 $\pm 0.05$ &  4.11 $\pm 0.07$ &  3.59 $\pm 0.05$ &  3.07 $\pm 0.07$ &  2.84 $\pm 0.05$ &  2.91 $\pm 0.08$ &  2.64 $\pm 0.06$\\
      ESO 381-47 &  1.40 $\pm 0.05$ &  1.65 $\pm 0.06$ &  4.93 $\pm 0.06$ &  4.33 $\pm 0.08$ &  3.01 $\pm 0.07$ &  2.60 $\pm 0.09$ &  2.84 $\pm 0.08$ &  2.31 $\pm 0.10$\\
         IC4 200 &  1.72 $\pm 0.06$ &  1.79 $\pm 0.05$ &  4.36 $\pm 0.07$ &  4.15 $\pm 0.06$ &  3.01 $\pm 0.08$ &  2.90 $\pm 0.06$ &  3.03 $\pm 0.08$ &  2.89 $\pm 0.07$\\
         IC 4889 &  1.89 $\pm 0.04$ &  1.88 $\pm 0.03$ &  4.18 $\pm 0.04$ &  3.98 $\pm 0.03$ &  3.22 $\pm 0.05$ &  3.04 $\pm 0.03$ &  3.05 $\pm 0.05$ &  2.87 $\pm 0.04$\\
        NGC 1490 &  1.60 $\pm 0.04$ &  1.51 $\pm 0.05$ &  5.50 $\pm 0.06$ &  5.13 $\pm 0.07$ &  3.17 $\pm 0.06$ &  2.85 $\pm 0.07$ &  3.29 $\pm 0.06$ &  3.14 $\pm 0.07$\\
        NGC 1947 &  1.84 $\pm 0.07$ &  1.93 $\pm 0.06$ &  4.02 $\pm 0.08$ &  3.87 $\pm 0.07$ &  2.91 $\pm 0.08$ &  2.63 $\pm 0.07$ &  2.75 $\pm 0.10$ &  2.45 $\pm 0.08$\\
        NGC 2434 &  1.73 $\pm 0.06$ &  1.84 $\pm 0.05$ &  5.00 $\pm 0.07$ &  4.20 $\pm 0.06$ &  2.98 $\pm 0.07$ &  2.65 $\pm 0.06$ &  2.92 $\pm 0.08$ &  2.60 $\pm 0.07$\\
        NGC 2904 &  1.58 $\pm 0.05$ &  1.57 $\pm 0.05$ &  5.64 $\pm 0.06$ &  4.92 $\pm 0.06$ &  3.18 $\pm 0.06$ &  2.87 $\pm 0.06$ &  3.32 $\pm 0.07$ &  2.94 $\pm 0.07$\\
        NGC 3108 &  1.50 $\pm 0.06$ &  1.64 $\pm 0.06$ &  5.27 $\pm 0.08$ &  4.81 $\pm 0.08$ &  3.27 $\pm 0.08$ &  3.00 $\pm 0.08$ &  3.27 $\pm 0.08$ &  2.94 $\pm 0.08$\\
\noalign{\smallskip} \noalign{\smallskip}
\multicolumn{9}{c}{\emph{WHT/ISIS spectroscopy}}\\
        NGC 0596 &  1.80 $\pm 0.04$ &  1.70 $\pm 0.05$ &  4.03 $\pm 0.05$ &  3.65 $\pm 0.06$ &  3.10 $\pm 0.06$ &  2.71 $\pm 0.07$ &  2.83 $\pm 0.07$ &  2.59 $\pm 0.08$\\
        NGC 0636 &  1.65 $\pm 0.05$ &  1.58 $\pm 0.06$ &  4.58 $\pm 0.06$ &  4.04 $\pm 0.08$ &  3.18 $ls Mgb\pm 0.07$ &  2.85 $\pm 0.08$ &  2.92 $\pm 0.08$ &  2.55 $\pm 0.09$\\
        NGC 1426 &  1.66 $\pm 0.04$ &  1.75 $\pm 0.05$ &  4.47 $\pm 0.06$ &  3.97 $\pm 0.07$ &  3.16 $\pm 0.06$ &  2.84 $\pm 0.07$ &  3.00 $\pm 0.07$ &  2.57 $\pm 0.08$\\
        NGC 1439 &  1.70 $\pm 0.05$ &  1.69 $\pm 0.06$ &  4.68 $\pm 0.06$ &  3.73 $\pm 0.08$ &  3.16 $\pm 0.07$ &  2.63 $\pm 0.09$ &  2.96 $\pm 0.07$ &  2.43 $\pm 0.10$\\
        NGC 2300 &  1.51 $\pm 0.06$ &  1.47 $\pm 0.06$ &  5.43 $\pm 0.08$ &  5.03 $\pm 0.08$ &  3.16 $\pm 0.09$ &  3.02 $\pm 0.09$ &  2.96 $\pm 0.10$ &  2.59 $\pm 0.10$\\
        NGC 2534 &  2.56 $\pm 0.06$ &  2.73 $\pm 0.09$ &  3.10 $\pm 0.09$ &  2.44 $\pm 0.12$ &  2.52 $\pm 0.10$ &  2.40 $\pm 0.13$ &  2.13 $\pm 0.11$ &  1.91 $\pm 0.15$\\
        NGC 2549 &  1.99 $\pm 0.05$ &  1.85 $\pm 0.05$ &  4.36 $\pm 0.06$ &  3.89 $\pm 0.07$ &  3.21 $\pm 0.07$ &  2.88 $\pm 0.08$ &  3.27 $\pm 0.08$ &  2.74 $\pm 0.09$\\
        NGC 2768 &  1.82 $\pm 0.07$ &  1.77 $\pm 0.07$ &  4.61 $\pm 0.09$ &  4.42 $\pm 0.10$ &  3.11 $\pm 0.10$ &  2.96 $\pm 0.10$ &  3.20 $\pm 0.11$ &  2.90 $\pm 0.11$\\
        NGC 2810 &  1.68 $\pm 0.06$ &  1.88 $\pm 0.08$ &  5.38 $\pm 0.08$ &  5.20 $\pm 0.11$ &  3.05 $\pm 0.08$ &  2.92 $\pm 0.12$ &  2.89 $\pm 0.09$ &  2.62 $\pm 0.13$\\
        NGC 3193 &  1.66 $\pm 0.05$ &  1.45 $\pm 0.07$ &  4.86 $\pm 0.07$ &  4.07 $\pm 0.09$ &  3.10 $\pm 0.08$ &  2.85 $\pm 0.10$ &  3.03 $\pm 0.09$ &  2.65 $\pm 0.12$\\
        NGC 3610 &  2.24 $\pm 0.03$ &  2.21 $\pm 0.03$ &  3.96 $\pm 0.04$ &  3.48 $\pm 0.04$ &  2.97 $\pm 0.05$ &  2.77 $\pm 0.05$ &  2.88 $\pm 0.06$ &  2.71 $\pm 0.05$\\
        NGC 3640 &  1.95 $\pm 0.06$ &  1.71 $\pm 0.06$ &  4.57 $\pm 0.08$ &  3.93 $\pm 0.08$ &  3.18 $\pm 0.08$ &  2.97 $\pm 0.09$ &  3.01 $\pm 0.09$ &  2.86 $\pm 0.10$\\
        NGC 4026 &  1.52 $\pm 0.07$ &  1.56 $\pm 0.07$ &  4.45 $\pm 0.10$ &  3.94 $\pm 0.09$ &  3.13 $\pm 0.10$ &  2.56 $\pm 0.10$ &  2.88 $\pm 0.11$ &  2.70 $\pm 0.11$\\
        NGC 4125 &  1.81 $\pm 0.04$ &  1.87 $\pm 0.04$ &  4.84 $\pm 0.05$ &  4.37 $\pm 0.05$ &  3.26 $\pm 0.06$ &  3.02 $\pm 0.05$ &  3.27 $\pm 0.06$ &  3.06 $\pm 0.06$\\
        NGC 4278 &  1.92 $\pm 0.05$ &  1.99 $\pm 0.05$ &  5.56 $\pm 0.07$ &  5.24 $\pm 0.07$ &  2.79 $\pm 0.07$ &  2.79 $\pm 0.07$ &  3.01 $\pm 0.08$ &  2.83 $\pm 0.08$\\
        NGC 4406 &  1.35 $\pm 0.06$ &  1.45 $\pm 0.06$ &  5.64 $\pm 0.07$ &  4.95 $\pm 0.07$ &  3.14 $\pm 0.07$ &  3.04 $\pm 0.07$ &  3.12 $\pm 0.09$ &  2.91 $\pm 0.09$\\
        NGC 4472 &  1.56 $\pm 0.06$ &  1.50 $\pm 0.05$ &  5.62 $\pm 0.08$ &  5.23 $\pm 0.07$ &  3.42 $\pm 0.08$ &  3.07 $\pm 0.07$ &  3.23 $\pm 0.10$ &  3.16 $\pm 0.08$\\
        NGC 5018 &  2.39 $\pm 0.06$ &  2.31 $\pm 0.07$ &  3.51 $\pm 0.09$ &  3.09 $\pm 0.10$ &  2.96 $\pm 0.09$ &  2.94 $\pm 0.10$ &  2.74 $\pm 0.10$ &  2.51 $\pm 0.12$\\
        NGC 5173 &  2.31 $\pm 0.04$ &  2.25 $\pm 0.05$ &  3.03 $\pm 0.05$ &  2.67 $\pm 0.06$ &  2.34 $\pm 0.06$ &  2.23 $\pm 0.07$ &  2.29 $\pm 0.07$ &  2.01 $\pm 0.08$\\
        NGC 5322 &  1.72 $\pm 0.06$ &  1.91 $\pm 0.06$ &  4.64 $\pm 0.08$ &  4.30 $\pm 0.08$ &  3.25 $\pm 0.08$ &  3.11 $\pm 0.08$ &  3.05 $\pm 0.09$ &  2.90 $\pm 0.09$\\
        NGC 5903 &  1.65 $\pm 0.06$ &  1.67 $\pm 0.07$ &  4.98 $\pm 0.08$ &  4.27 $\pm 0.09$ &  3.01 $\pm 0.08$ &  2.63 $\pm 0.10$ &  2.86 $\pm 0.10$ &  2.52 $\pm 0.11$\\
        NGC 7332 &  2.23 $\pm 0.03$ &  2.04 $\pm 0.04$ &  3.97 $\pm 0.04$ &  3.40 $\pm 0.05$ &  2.99 $\pm 0.05$ &  2.68 $\pm 0.06$ &  2.94 $\pm 0.05$ &  2.50 $\pm 0.06$\\
        NGC 7457 &  2.37 $\pm 0.05$ &  2.09 $\pm 0.06$ &  3.10 $\pm 0.07$ &  2.90 $\pm 0.08$ &  2.72 $\pm 0.08$ &  2.39 $\pm 0.09$ &  2.58 $\pm 0.09$ &  2.36 $\pm 0.10$\\
        NGC 7585 &  2.25 $\pm 0.06$ &  2.04 $\pm 0.07$ &  3.94 $\pm 0.08$ &  3.59 $\pm 0.10$ &  2.99 $\pm 0.09$ &  2.79 $\pm 0.11$ &  2.84 $\pm 0.10$ &  2.32 $\pm 0.13$\\
        NGC 7600 &  2.37 $\pm 0.05$ &  2.06 $\pm 0.08$ &  3.40 $\pm 0.08$ &  3.25 $\pm 0.11$ &  2.96 $\pm 0.09$ &  2.73 $\pm 0.13$ &  2.59 $\pm 0.10$ &  2.23 $\pm 0.14$\\
        NGC 7619 &  1.32 $\pm 0.05$ &  1.42 $\pm 0.06$ &  5.72 $\pm 0.08$ &  5.17 $\pm 0.09$ &  3.44 $\pm 0.09$ &  3.10 $\pm 0.10$ &  3.12 $\pm 0.09$ &  2.86 $\pm 0.11$\\
        NGC 7626 &  1.14 $\pm 0.05$ &  1.23 $\pm 0.07$ &  5.73 $\pm 0.07$ &  4.89 $\pm 0.10$ &  3.21 $\pm 0.09$ &  2.79 $\pm 0.11$ &  3.12 $\pm 0.09$ &  2.80 $\pm 0.12$\\
\noalign{\smallskip}
\hline
\end{tabular}
\end{center}
The four indices used in this paper for all galaxies and apertures. The error is 1 $\sigma$ estimated from the spectrum noise only. The H$\beta$ index is already corrected for ionised-gas emission (see Table \ref{t3}).
\end{table*}

\begin{table*}
\begin{center}
\caption{Derived SSP-equivalent stellar population parameters}
\label{t5}
\begin{tabular}{lcccccc}
\hline
\hline
\noalign{\smallskip}
\multicolumn{1}{c}{galaxy}      & \multicolumn{2}{c}{log$_{10}$ \agessp\ (Gyr)} & \multicolumn{2}{c}{\metalssp} & \multicolumn{2}{c}{\abundssp} \\
            & $R_e$/16 & $R_e$/2           & $R_e$/16 & $R_e$/2        & $R_e$/16 & $R_e$/2 \\
\noalign{\smallskip} \hline \noalign{\smallskip}
\noalign{\smallskip} \noalign{\smallskip}
\multicolumn{7}{c}{\emph{NTT/EMMI spectroscopy}}\\
ESO 092-21 &  0.72  [ 0.64, 0.83] &  0.77  [ 0.64, 0.80] &  0.24  [ 0.18, 0.30] & -0.13  [-0.17,-0.08] &  0.11  [ 0.09, 0.13] &  0.08  [ 0.06, 0.10] \\
      ESO 140-31 &  0.74  [ 0.71, 0.83] &  0.88  [ 0.85, 0.91] &  0.22  [ 0.16, 0.23] & -0.06  [-0.08,-0.03] &  0.07  [ 0.06, 0.08] &  0.03  [ 0.03, 0.04] \\
      ESO 381-47 &  1.11  [ 1.07, 1.16] &  0.95  [ 0.94, 1.03] &  0.18  [ 0.15, 0.21] & -0.02  [-0.05, 0.01] &  0.18  [ 0.17, 0.19] &  0.21  [ 0.20, 0.23] \\
         IC 4200 &  0.82  [ 0.75, 0.91] &  0.78  [ 0.73, 0.82] &  0.24  [ 0.21, 0.29] &  0.18  [ 0.15, 0.21] &  0.11  [ 0.10, 0.12] &  0.10  [ 0.09, 0.11] \\
         IC 4889 &  0.70  [ 0.66, 0.72] &  0.74  [ 0.73, 0.76] &  0.30  [ 0.28, 0.31] &  0.18  [ 0.17, 0.18] &  0.06  [ 0.06, 0.07] &  0.06  [ 0.06, 0.06] \\
        NGC 1490 &  0.90  [ 0.67, 0.90] &  0.90  [ 0.78, 0.96] &  0.45  [ 0.45, 0.53] &  0.36  [ 0.33, 0.38] &  0.22  [ 0.20, 0.22] &  0.20  [ 0.20, 0.21] \\
        NGC 1947 &  0.76  [ 0.73, 0.86] &  0.77  [ 0.72, 0.81] &  0.13  [ 0.09, 0.18] &  0.01  [-0.02, 0.06] &  0.09  [ 0.07, 0.10] &  0.14  [ 0.13, 0.15] \\
        NGC 2434 &  0.73  [ 0.67, 0.78] &  0.75  [ 0.72, 0.81] &  0.40  [ 0.38, 0.42] &  0.15  [ 0.12, 0.18] &  0.22  [ 0.21, 0.23] &  0.18  [ 0.17, 0.19] \\
        NGC 2904 &  0.89  [ 0.67, 0.89] &  0.90  [ 0.90, 0.97] &  0.48  [ 0.46, 0.55] &  0.31  [ 0.28, 0.32] &  0.23  [ 0.22, 0.24] &  0.20  [ 0.20, 0.21] \\
        NGC 3108 &  0.90  [ 0.90, 1.02] &  0.91  [ 0.80, 0.96] &  0.43  [ 0.40, 0.44] &  0.30  [ 0.27, 0.33] &  0.17  [ 0.16, 0.18] &  0.18  [ 0.17, 0.19] \\
\noalign{\smallskip} \noalign{\smallskip}
\multicolumn{7}{c}{\emph{WHT/ISIS spectroscopy}}\\
        NGC 0596 &  0.83  [ 0.81, 0.88] &  0.97  [ 0.94, 1.02] &  0.14  [ 0.11, 0.17] & -0.14  [-0.15,-0.10] &  0.06  [ 0.05, 0.07] &  0.05  [ 0.04, 0.06] \\
        NGC 0636 &  0.93  [ 0.87, 0.96] &  1.05  [ 1.01, 1.10] &  0.27  [ 0.22, 0.27] & -0.07  [-0.09,-0.04] &  0.12  [ 0.11, 0.13] &  0.10  [ 0.08, 0.11] \\
        NGC 1426 &  0.92  [ 0.84, 0.95] &  0.91  [ 0.88, 0.94] &  0.25  [ 0.22, 0.26] & -0.02  [-0.03, 0.03] &  0.10  [ 0.09, 0.11] &  0.10  [ 0.09, 0.10] \\
        NGC 1439 &  0.76  [ 0.73, 0.85] &  1.00  [ 0.96, 1.05] &  0.30  [ 0.30, 0.34] & -0.15  [-0.18,-0.11] &  0.14  [ 0.13, 0.15] &  0.10  [ 0.09, 0.12] \\
        NGC 2300 &  0.98  [ 0.77, 1.02] &  1.07  [ 0.99, 1.11] &  0.40  [ 0.37, 0.43] &  0.20  [ 0.17, 0.24] &  0.23  [ 0.22, 0.24] &  0.22  [ 0.20, 0.23] \\
        NGC 2534 &  0.40  [ 0.35, 0.44] &  0.38  [ 0.35, 0.40] &  0.07  [ 0.01, 0.10] & -0.23  [-0.30,-0.15] &  0.12  [ 0.10, 0.15] &  0.00  [-0.05, 0.03] \\
        NGC 2549 &  0.51  [ 0.42, 0.54] &  0.82  [ 0.76, 0.85] &  0.45  [ 0.42, 0.48] &  0.10  [ 0.06, 0.13] &  0.12  [ 0.11, 0.13] &  0.07  [ 0.05, 0.08] \\
        NGC 2768 &  0.63  [ 0.53, 0.74] &  0.74  [ 0.69, 0.86] &  0.39  [ 0.35, 0.43] &  0.27  [ 0.21, 0.30] &  0.14  [ 0.11, 0.15] &  0.14  [ 0.12, 0.15] \\
        NGC 2810 &  0.73  [ 0.58, 0.76] &  0.58  [ 0.49, 0.69] &  0.48  [ 0.45, 0.49] &  0.43  [ 0.40, 0.52] &  0.26  [ 0.25, 0.28] &  0.31  [ 0.29, 0.34] \\
        NGC 3193 &  0.77  [ 0.72, 0.84] &  1.15  [ 1.09, 1.23] &  0.35  [ 0.32, 0.37] & -0.10  [-0.14,-0.06] &  0.17  [ 0.16, 0.18] &  0.09  [ 0.08, 0.11] \\
        NGC 3610 &  0.37  [ 0.37, 0.40] &  0.48  [ 0.48, 0.49] &  0.39  [ 0.38, 0.41] &  0.14  [ 0.13, 0.18] &  0.14  [ 0.13, 0.14] &  0.07  [ 0.06, 0.07] \\
        NGC 3640 &  0.53  [ 0.48, 0.59] &  0.92  [ 0.87, 0.96] &  0.42  [ 0.39, 0.46] & -0.01  [-0.02, 0.08] &  0.17  [ 0.15, 0.18] &  0.04  [ 0.03, 0.06] \\
        NGC 4026 &  1.05  [ 0.99, 1.12] &  1.10  [ 1.01, 1.15] &  0.11  [ 0.05, 0.17] & -0.13  [-0.16,-0.08] &  0.09  [ 0.08, 0.11] &  0.11  [ 0.09, 0.13] \\
        NGC 4125 &  0.58  [ 0.53, 0.64] &  0.67  [ 0.64, 0.68] &  0.46  [ 0.46, 0.49] &  0.33  [ 0.32, 0.35] &  0.16  [ 0.15, 0.17] &  0.12  [ 0.11, 0.12] \\
        NGC 4278 &  0.48  [ 0.48, 0.64] &  0.34  [ 0.31, 0.38] &  0.60  [ 0.49, 0.67] &  0.59  [ 0.54, 0.63] &  0.35  [ 0.34, 0.38] &  0.35  [ 0.34, 0.36] \\
        NGC 4406 &  1.04  [ 1.00, 1.09] &  1.05  [ 1.01, 1.11] &  0.40  [ 0.39, 0.43] &  0.23  [ 0.20, 0.28] &  0.23  [ 0.23, 0.24] &  0.18  [ 0.17, 0.19] \\
        NGC 4472 &  0.90  [ 0.72, 0.98] &  0.90  [ 0.90, 0.99] &  0.47  [ 0.47, 0.52] &  0.39  [ 0.38, 0.41] &  0.20  [ 0.19, 0.21] &  0.19  [ 0.19, 0.20] \\
        NGC 5018 &  0.36  [ 0.33, 0.40] &  0.47  [ 0.47, 0.52] &  0.31  [ 0.25, 0.34] &  0.01  [-0.04, 0.05] &  0.07  [ 0.05, 0.09] & -0.02  [-0.05, 0.00] \\
        NGC 5173 &  0.51  [ 0.49, 0.60] &  0.55  [ 0.55, 0.72] & -0.15  [-0.17,-0.11] & -0.37  [-0.39,-0.33] &  0.07  [ 0.06, 0.09] &  0.03  [ 0.02, 0.05] \\
        NGC 5322 &  0.76  [ 0.71, 0.84] &  0.69  [ 0.59, 0.72] &  0.34  [ 0.30, 0.36] &  0.31  [ 0.28, 0.34] &  0.12  [ 0.11, 0.13] &  0.11  [ 0.10, 0.13] \\
        NGC 5903 &  0.77  [ 0.74, 0.92] &  0.93  [ 0.90, 1.00] &  0.34  [ 0.31, 0.37] & -0.01  [-0.04, 0.06] &  0.21  [ 0.19, 0.22] &  0.17  [ 0.16, 0.20] \\
        NGC 7332 &  0.39  [ 0.38, 0.42] &  0.72  [ 0.70, 0.73] &  0.40  [ 0.39, 0.41] & -0.05  [-0.07,-0.04] &  0.13  [ 0.13, 0.14] &  0.04  [ 0.04, 0.06] \\
        NGC 7457 &  0.46  [ 0.46, 0.48] &  0.72  [ 0.71, 0.77] &  0.08  [ 0.02, 0.12] & -0.26  [-0.28,-0.21] &  0.02  [ 0.01, 0.04] &  0.01  [-0.01, 0.03] \\
        NGC 7585 &  0.38  [ 0.35, 0.42] &  0.72  [ 0.65, 0.75] &  0.38  [ 0.34, 0.41] &  0.01  [-0.05, 0.02] &  0.13  [ 0.12, 0.15] &  0.09  [ 0.07, 0.12] \\
        NGC 7600 &  0.46  [ 0.40, 0.47] &  0.73  [ 0.69, 0.79] &  0.21  [ 0.17, 0.26] & -0.12  [-0.16,-0.07] &  0.06  [ 0.05, 0.08] &  0.04  [ 0.02, 0.07] \\
        NGC 7619 &  1.12  [ 1.04, 1.13] &  1.05  [ 0.99, 1.11] &  0.40  [ 0.38, 0.43] &  0.27  [ 0.24, 0.32] &  0.21  [ 0.20, 0.22] &  0.20  [ 0.20, 0.23] \\
        NGC 7626 &  1.26  [ 1.20, 1.35] &  1.28  [ 1.18, 1.45] &  0.31  [ 0.29, 0.34] & -0.00  [-0.02, 0.07] &  0.24  [ 0.23, 0.25] &  0.20  [ 0.18, 0.21] \\
\noalign{\smallskip}
\hline
\end{tabular}
\end{center}
Stellar parameters estimated from the index values in Table \ref{t4} for all galaxies and apertures. For each parameter we give the peak value and the parameter range containing 68\% of the probability distribution (see Sec.\ref{lick2ssp}).
\end{table*}

\begin{appendix}
\section{Calibration of the line-strength indices to the Lick/IDS system}
\label{Lick/IDScal}

\begin{figure*}
\begin{center}
\includegraphics[width=17cm]{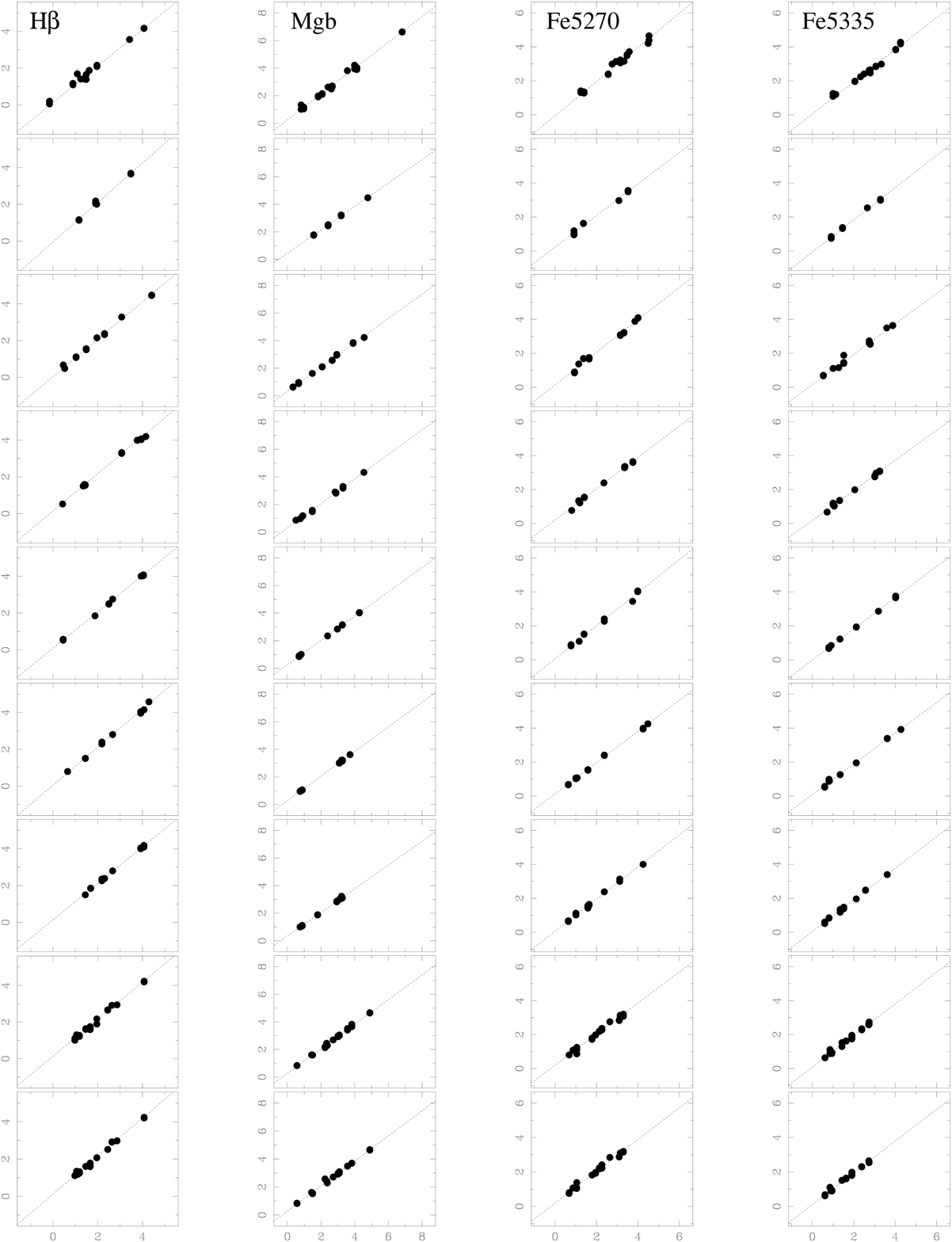}
\caption{Calibration to the Lick/IDS system for all observing nights, one night per row with rows 1-5 and 6-9 corresponding to NTT/EMMI and WHT/ISIS runs as in Table \ref{calick}. For each index the points on the plane ($I_{\rm ref},I_{\rm meas}$) are showed together with the best linear fit whose coefficients are reported in Table \ref{calick}. Index values are in \AA. The fitting was performed weighting points according to their errors in both coordinates. Errors are too small to be shown in the plots.}
\label{lickcalib}
\end{center}
\end{figure*}

\begin{table}
\begin{center}
\caption{INDO-US to MILES transformation}
\label{I2M}
\begin{tabular}{c c c}
\hline
\hline
\noalign{\smallskip}
Index & $m$ & $q$ \\
\noalign{\smallskip} \hline \noalign{\smallskip}
H$\beta$    & 1.050 &    0.047 \\
Mg$b$       & 0.898 &    0.257 \\
Fe5270      & 0.954 &    0.122 \\
Fe5335      & 0.906 &    0.062 \\
\noalign{\smallskip}
\hline
\end{tabular}
\end{center}
$m$ and $q$ are respectively slope and intercept of the transformation $I_{\rm INDO-US}=m\times{I_{\rm MILES}}+q$.
\end{table}

\begin{table}
\begin{center}
\caption{MILES to Lick/IDS transformation}
\label{M2L}
\begin{tabular}{c c c}
\hline
\hline
\noalign{\smallskip}
Index & $m$ & $q$ \\
\noalign{\smallskip} \hline \noalign{\smallskip}
H$\beta$    & 0.976 &    0.154 \\
Mg$b$       & 0.893 &    0.220 \\
Fe5270      & 0.901 &    0.131 \\
Fe5335      & 0.927 &    0.106 \\
\noalign{\smallskip}
\hline
\end{tabular}
\end{center}
$m$ and $q$ are respectively slope and intercept of the transformation $I_{\rm MILES}=m\times{I_{\rm Lick/IDS}}+q$.
\end{table}

\begin{table*}
\begin{center}
\caption{Calibration to Lick/IDS system}
\label{calick}
\begin{tabular}{lrrrrrrrrl}
\hline
\hline
\noalign{\smallskip}
Night & \multicolumn{2}{c}{H$\beta$} & \multicolumn{2}{c}{Mg$b$} & \multicolumn{2}{c}{Fe5270} & \multicolumn{2}{c}{Fe5335} & \multicolumn{1}{c}{Stars} \\
      & \multicolumn{1}{c}{$m$} & \multicolumn{1}{c}{$q$} & \multicolumn{1}{c}{$m$} & \multicolumn{1}{c}{$q$} & \multicolumn{1}{c}{$m$} & \multicolumn{1}{c}{$q$} & \multicolumn{1}{c}{$m$} & \multicolumn{1}{c}{$q$} & \multicolumn{1}{c}{(HD name)} \\
\noalign{\smallskip} \hline \noalign{\smallskip}
1 & 0.992 &  0.126 & 0.937 &  0.245 & 1.006 & -0.058 & 0.946 &  0.045 & 114113*, 124850**, 125454*, 126218, 131430, \\
  &       &        &       &        &       &        &       &        & 131977, 137052*, 138716*, 138905*, 139446, \\
  &       &        &       &        &       &        &       &        & 1142198* \\
2 & 1.072 & -0.046 & 0.848 &  0.453 & 0.928 &  0.237 & 0.950 & -0.062 & 025457*, 126681**, 129978*, 130322 \\
3 & 1.016 &  0.042 & 0.882 &  0.315 & 0.957 &  0.146 & 0.895 &  0.153 & 029065, 029574*, 049933, 050778, 091889**, \\
  &       &        &       &        &       &        &       &        & 125454*, 126053, 126218 \\
4 & 1.013 &  0.147 & 0.890 &  0.284 & 0.923 &  0.204 & 0.887 &  0.178 & 037792, 041312, 091889**, 114642**, 114946**, \\
  &       &        &       &        &       &        &       &        & 120452*, 121299**, 122106 \\
5 & 0.988 &  0.093 & 0.880 &  0.263 & 0.973 &  0.060 & 0.920 & -0.017 & 033256, 065953, 076151**, 078558**, 078738, \\
  &       &        &       &        &       &        &       &        & 082734  \\
6 & 1.033 &  0.053 & 0.892 &  0.276 & 0.920 &  0.125 & 0.902 &  0.088 & 069267, 073471, 076151, 089995, 097916, \\
  &       &        &       &        &       &        &       &        & 114606, 119288 \\
7 & 0.990 &  0.137 & 0.858 &  0.344 & 0.946 &  0.065 & 0.933 &  0.041 & 073471, 076151, 078732, 089995, 097916, \\
  &       &        &       &        &       &        &       &        & 114606, 126053 \\
8 & 1.009 &  0.135 & 0.890 &  0.275 & 0.913 &  0.208 & 0.932 &  0.090 & 010975, 013783, 014221, 015596, 051440, \\
  &       &        &       &        &       &        &       &        & 062301, 065583, 072324, 191046, 199191, \\
  &       &        &       &        &       &        &       &        & 201889, 201891\\
9 & 1.019 &  0.114 & 0.895 &  0.281 & 0.923 &  0.190 & 0.921 &  0.107 & as night 8\\
\noalign{\smallskip}
\hline
\end{tabular}
\end{center}
Slope $m$ and intercept $q$ of the linear relation between measured indices and Lick/IDS-reference values for all observing nights. Nights 1 to 5 are NTT/EMMI runs, 6 to 9 are WHT/ISIS runs (see Table \ref{longslit}). The coefficients correspond to the relation $i_{meas}=m\times i_{ref}+q$. The last column gives the HD name of the calibration stars observed during each night. An asterisk marks stars belonging to the INDO-US library only. Two asterisks mark stars belonging to both MILES and INDO-US library. All other stars belong to the MILES library only.
\end{table*}

Our stellar population analysis relies on the comparison of galaxy line-strength indices to the values predicted by BC03 SSP models (adequately modified to take into account non solar [E/Fe]). Model indices are given onto the Lick/IDS system, which is defined by the instrumental conditions under which stars in the Lick/IDS library were observed (Worthey et al.\ 1994). Therefore, any comparison of a galaxy spectral indices to these models makes sense only if the measured indices are themselves onto the Lick/IDS system. A practical way to satisfy this requirement is to observe a sample of stars that already have indices on the Lick/IDS system with the same instrumental setup used for the galaxy. We can then measure the spectral indices of these stars from our spectra and calculate the transformation necessary to make them match their Lick/IDS-reference values. The same transformation will then be applied to the indices measured from the galaxy spectrum. In our case we perform the calibration for each observing night listed in Table \ref{longslit}.

We observe stars belonging to the MILES library (S\'{a}nchez-Bl\'{a}zquez et al.\ 2006) and/or to the INDO-US library (Valdes et al.\ 2004). The stars observed during each observing nights are listed in Table \ref{calick}. In order to build a set of Lick/IDS-reference indices we measure index values from the MILES and INDO-US original spectra broadened to the Lick/IDS resolution. We then use the 8 observed stars in common between MILES and INDO-US library to calculate linear transformations that bring the INDO-US indices onto the MILES system (see Table \ref{I2M}), and apply such transformations to all 17 INDO-US stars.  We finally bring all indices onto the Lick/IDS system with the MILES-to-Lick/IDS transformations provided by P. S\'{a}nchez-Bl\'{a}zquez (see Table \ref{M2L}).

Lick/IDS-reference index values are compared to the ones measured from the observed stellar spectra broadened to the Lick/IDS resolution. A linear transformation between the two is computed and reported in Table\ref{calick} for each night and each of the indices used in this work. The transformations in Table \ref{calick} are finally applied to galaxy indices before performing the comparison to BC03 models in order to obtain stellar population parameters. Fig.\ref{lickcalib} shows such linear relations.

\end{appendix}

\end{document}